\documentclass[a4wide,11pt,oneside]{book} 

\input{add/settings.tex}

\usepackage{color,xcolor}
\usepackage[export]{adjustbox}

\definecolor{amethyst}{rgb}{0.6, 0.4, 0.8}
\definecolor{green}{rgb}{0.55, 0.71, 0.0}
\definecolor{apricot}{rgb}{0.98, 0.81, 0.69}
\definecolor{auburn}{rgb}{0.43, 0.21, 0.1}
\definecolor{babyblueeyes}{rgb}{0.63, 0.79, 0.95}
\definecolor{bittersweet}{rgb}{1.0, 0.44, 0.37}
\definecolor{officegreen}{rgb}{0.0, 0.5, 0.0}
\definecolor{darkcandyapplered}{rgb}{0.64, 0.0, 0.0}
\definecolor{blue(munsell)}{rgb}{0.0, 0.5, 0.69}

\newcommand{\strike}{\bgroup\markoverwith{\textcolor{red}{\rule[0.5ex]{2pt}{0.4pt}}}\ULon}
\usepackage{soul} 

\begin{document}

\noindent
\textbf{{\Large Cosmology with Very-High-Energy Gamma Rays}}

\bigskip\noindent
To appear in `Advances in Very High Energy Astrophysics', Mukherjee \& Zanin, World Scientific (2022) \url{https://www.worldscientific.com/worldscibooks/10.1142/11141}

\bigskip\bigskip
\noindent
Elisa Pueschel$^1$ and Jonathan Biteau$^2$

\medskip
\noindent
$^1$ Deutsches Elektronen-Synchrotron, D-15738 Zeuthen, Germany \url{elisa.pueschel@desy.de}\\
$^2$ Universit\'e Paris-Saclay, CNRS/IN2P3, IJCLab, 91405 Orsay, France \url{biteau@in2p3.fr} \\

\newpage
\setcounter{chapter}{7}

\section*{Introduction}

Modern cosmology depicts the history of the universe as starting from a hot initial phase, filled with particles and antiparticles and expanding to eventually give rise to large-scale structures. While the first moments of the universe, near the Planck scale at $t_{\rm Pl} = 5.4 \times 10^{-44}\:$s, remain elusive on theoretical grounds, modern theories successfully describe the large-scale components and composition of the universe through a succession of epochs: inflation providing the flatness and causal connection of our universe; abrupt transitions describing the symmetry breaking of the electroweak and strong forces; decoupling of matter and radiation giving rise to the cosmic microwave background (CMB); the dark ages ending with the emergence of the first astrophysical sources of reionization; and finally the formation of structures and galaxies.

Observations constraining Big Bang Nucleosynthesis, the cosmic distance ladder, the spectrum and anisotropies of the CMB, and large-scale structures have been instrumental over the past fifty years in forming the modern vision of the universe's history, yielding the current $\Lambda$CDM model largely dominated by cold dark matter (CDM) and dark energy ($\Lambda$). In this chapter, we discuss the contributions of gamma-ray astronomy at TeV energies to our understanding of the visible content and structure of the universe. We start from the present epoch with the second most intense  electromagnetic background field after the CMB: the extragalactic background light (EBL). The EBL is composed of all the light emitted by stars and galaxies since the beginning of reionization, including light absorbed and re-emitted by dust. As such, the EBL traces the history of radiating matter in the universe. We then further dive into the large voids of the universe to study the large-scale magnetic fields that should permeate them. These fields could originate from the onset of structure formation or early phase transitions, bringing us back to the infancy of the universe. We conclude by looking back to the elusive Planck time scale, where the standard models of cosmology and particle physics are no longer applicable. Observations with current-generation gamma-ray astronomy experiments have now started to scratch the surface of cosmology, as we will show in this chapter.

\section{The extragalactic background light}

The study of the EBL is intertwined with an old scientific riddle, introduced by Digges in his English translation of Copernicus' work (1576), further discussed by astronomers such as Kepler, Huygens, Halley, to be finally formulated in its mathematical version by de Ch\'eseaux in 1744. 

Let us divide a static, homogeneous and infinite universe in concentric shells of radius $r$ and width $dr$. Each layer contains $4\pi r^2dr\rho_{*}$ galaxies, with $\rho_{*}$ their number density, and each galaxy emits a flux $L_{*}/4\pi r^2$, with $L_{*}$ its individual luminosity. The flux reaching the observer from each shell is then a constant, $L_{*}\rho_{*}dr$, and the sum from multiple shells diverges when integrated over sufficiently great distances, resulting in an infinitely bright sky. This is known as the ``dark night sky paradox'' or Olbers' paradox, from the amateur astronomer who popularized the riddle in the 19th century \citep{1991ApJ...367..399W}. Both de Ch\'eseaux and Olbers proposed an absorbing medium as a solution, but Herschel and Kelvin argued that absorption would inescapably lead to re-emission at longer wavelengths. The elucidation of the problem shook the premise of an infinite and immutable universe. Instead, light emission has a history which, from the epoch of reionization to the present epoch, is summarized in the starlight that composes the EBL.

\subsection{Production and evolution: modelling the EBL}
\label{sec:EBLmodel}
\begin{figure}[t]
\centering
\includegraphics[width=0.8\textwidth]{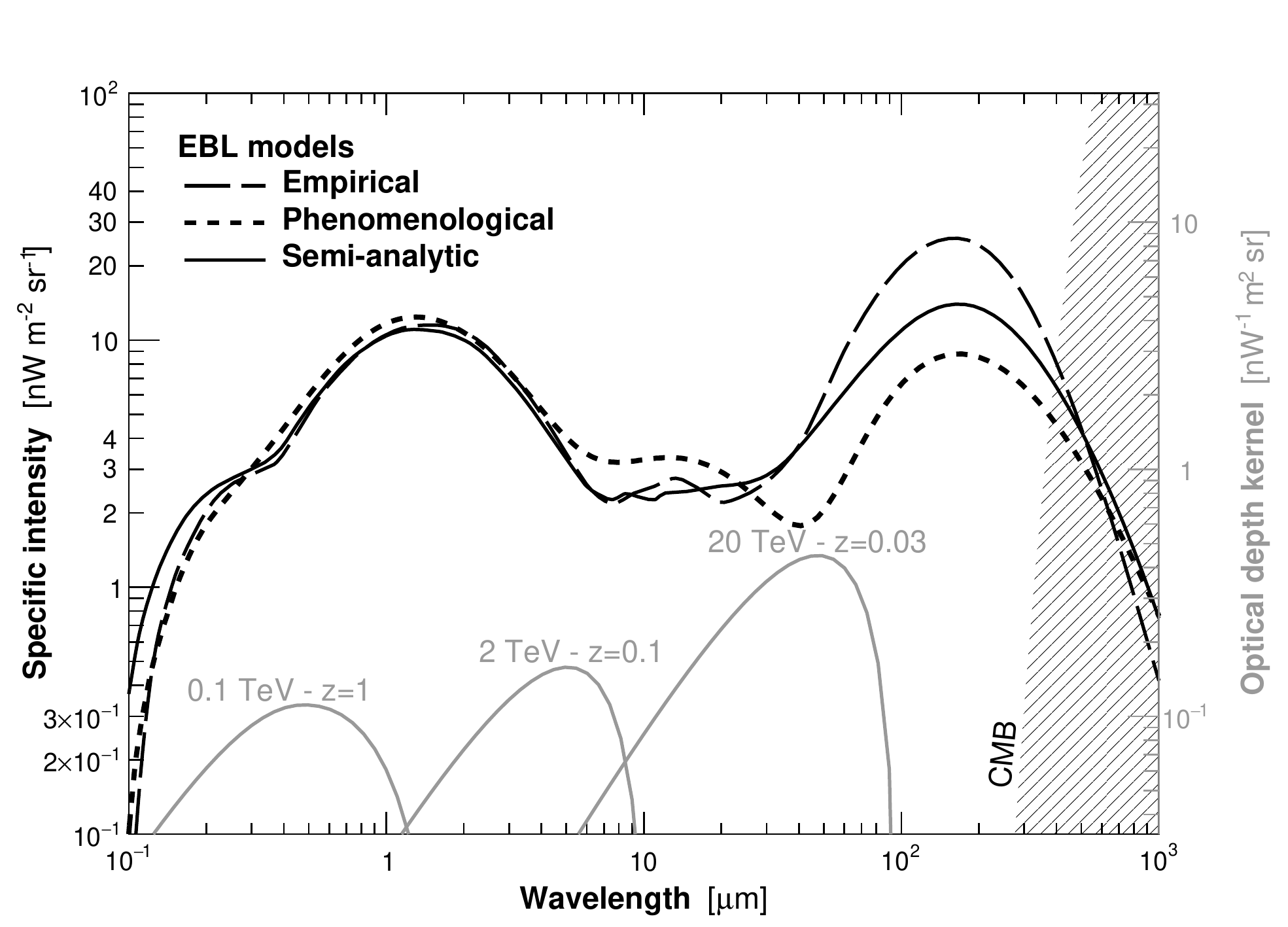}
\caption{Spectrum of the EBL at $z=0$, following the empirical model of \citet{2011MNRAS.410.2556D}, the phenomenological model of \citet{2010ApJ...712..238F} and the semi-analytical model of \citet{2012MNRAS.422.3189G}. The spectrum of the CMB, peaking around 1000\:nW\:m$^{-2}$\:sr$^{-1}$, is shown as a dashed area. Grey curves, denoted as ``optical depth kernels'' on the right-hand side axis, display the integral over redshift and angle of interaction of the pair production cross-section and evolution factors. The integral of the product of these functions and of the EBL intensity results in the gamma-ray optical depth, provided here at three characteristic energies and redshifts.}
\label{ch7:fig:EBLspectrum}
\end{figure}

The EBL intensity and its shape as a function of wavelength are dictated by the cosmic star formation history (CSFH). Most of the direct and reprocessed emission from stars lies in the spectral range 0.1\:$\mu$m-1000\:$\mu$m: limited at its lower end by the Lyman limit (0.0912\:$\mu$m) and dominated at its higher end by the CMB. 
About half of the starlight emitted in the ultraviolet (UV, 0.1-0.3\:$\mu$m), optical (0.3-0.8\:$\mu$m) and near-infrared (NIR, 0.8-3\:$\mu$m) bands escapes from galaxies, while the other half is absorbed by dust and reprocessed into the far-infrared band (FIR, 50-1000\:$\mu$m). Accretion of matter onto supermassive black holes of active galactic nuclei (AGN) constitutes a secondary source of UV and optical photons, about half of which are absorbed and re-radiated in the mid-infrared band (MIR, 3-50\:$\mu$m) by the hot dusty torii surrounding the accretion disks. Integrating the UV-FIR emission in time over the cosmic ages results in the broad-band EBL spectrum at $z=0$ presented in Fig.~\ref{ch7:fig:EBLspectrum}. It features two broad components: the cosmic optical background (COB, 0.1-8\:$\mu$m), composed of photons which escaped their environment without attenuation, and the cosmic infrared background (CIB, 8-1000\:$\mu$m), composed of photons absorbed and re-radiated by dust. Because about half the light emitted by the sources of the EBL escapes directly from its environment and half is reprocessed, the bolometric intensities of the COB and of the CIB end up being comparable, both being compatible with 30\:nW\:m$^{-2}$\:sr$^{-1}$, corresponding to a total energy budget for the EBL at the level of 6-7\% that of the CMB. Three state-of-the-art models are compared in Fig.~\ref{ch7:fig:EBLspectrum}, each representing a distinct sub-class: empirical, phenomenological, and semi-analytic models.

Deep-field observations from UV-FIR satellites such as GALEX, Hubble, Spitzer, and Herschel have been key in deriving the luminosity functions of various galaxy types, down to sufficiently low luminosities that the integrated galaxy light converges at most wavelengths~\citep{2016ApJ...827..108D}. These luminosity functions and associated observations are an important input to EBL models.

Empirical models (e.g. \citealp{2011MNRAS.410.2556D,2017A&A...603A..34F}) exploit galaxy counts and inferred luminosity functions to cumulate average UV-to-FIR template spectra from various sub-types of galaxies. 
Phenomenological models (e.g. \citealp{2010ApJ...712..238F, 2018MNRAS.474..898A}) additionally aim to reproduce the CSFH and evolution of metallicity. Combining constraints on the cosmic star-formation rate with the emission expected from each star, obtained e.g.\ from stellar population synthesis models for a given initial mass function, yields the direct emission contributing to the COB. The CIB is modelled assuming a fraction of reprocessed light and using template emission spectra for AGN and dust. These templates use high-quality spectral observations of galaxies, and include emission from amorphous or polycylic aromatic hydrocarbons, resulting in the bump around ${\sim}15\:\mu$m visible in Fig.~\ref{ch7:fig:EBLspectrum}. 
Finally, semi-analytical models (e.g. \citealp{2012MNRAS.422.3189G}) additionally aim at tracking accretion throughout cosmic ages by exploiting N-body simulations of dark matter halo mergers. Such models can include baryonic physics, in particular AGN and supernova feedback, which affects the star-formation rate of their host galaxies. 

All three types of models fundamentally aim at reproducing the emission from known, observed galaxies and as such reproduce in a fairly consistent manner the integrated galaxy light inferred from galaxy counts. This consistency is illustrated by a good match of the models around the peak of the COB at 1\:$\mu$m. While some uncertainties remain in the UV range (see \citealp{2019MNRAS.484.4174K} and references therein), the clearest mismatch between the models is observed at mid- and far-infrared wavelengths, where the amount of reprocessing by dust in the environment of the sources plays a key role. It should be noted that as current models aim at reproducing galaxy counts, they necessarily miss any unresolved population, such as primordial stars, and any truly diffuse component.

\subsection{VHE observables}

The very-high-energy (VHE, $E_\gamma > 100\:$GeV) band opens a unique window on the density of EBL photons integrated over the line of sight, be they from known galaxies or of truly diffuse origin (see e.g. \citealp{2021Univ....7..146F} for a recent review). Gamma rays from extragalactic sources can interact with EBL photons through the production of electron-positron pairs \citep{REF::NIKISHOV::JETP1962,1967PhRv..155.1408G,1967PhRv..155.1404G}. For this process to occur, the product of the gamma-ray energy, $E_\gamma'$, and of the EBL photon energy, $\epsilon'$, as measured in the cosmological comoving frame, must satisfy a threshold condition imposed by kinematics, that is:
\begin{equation}
\label{ch7:eq:EBLthreshold}
E_\gamma' \epsilon' \geq \left(m_{\rm e} c^2\right)^2
\end{equation}
where $m_{\rm e} c^2$ is the rest energy of the electron/positron.

The gamma-ray optical depth, $\tau(E_\gamma,z_0)$, with $E_\gamma$ the gamma-ray energy in the observer's frame and $z_0$ the redshift of the source, quantifies the number of interactions on the line of sight. As indicated in Eq.~\eqref{ch7:eq:opticaldepth}, $\tau(E_\gamma,z_0)$ is the product of the EBL photon density, $\frac{\partial n}{\partial \epsilon}$ (units of cm$^{-3}$\:eV$^{-1}$), and of the differential cross section for pair production, $\sigma_{\gamma \gamma}$ (units of cm$^{2}$), integrated over the distance between the observer and the source, $\frac{\partial L}{\partial z}$ (units of cm), over the energy of the EBL photons, $\epsilon'$ (units of eV), and over the angle, $\theta$, between the momenta of the gamma-ray and target EBL photons. 

\begin{equation}
\label{ch7:eq:opticaldepth}
\begin{split}
\tau{(E_\gamma, z_0)} = & \int^{z_0}_0 dz \frac{\partial L}{\partial z}(z) \int^{\infty}_0 d\epsilon' \frac{\partial n}{\partial \epsilon'}(\epsilon', z) \\
                   & \int^{-1}_1 d\cos \theta \frac{1- \cos \theta}{2} \sigma_{\gamma \gamma}\left(E_\gamma \times (1+z),\epsilon,\cos \theta\right)
\end{split}
\end{equation}
Using $\mu=\cos\theta$, the term $(1-\mu)/2$ in the integral normalizes the cross section so that if it were isotropic ($\sigma_{\gamma\gamma}$ independent from $\theta$), one would obtain $\int_{-1}^{1} d\mu \frac{1-\mu}{2}\sigma_{\gamma\gamma} = \sigma_{\gamma\gamma}$.

The distance element in a flat $\Lambda$CDM cosmology, with $H_0$ the Hubble constant, $\Omega_{\Lambda}$ the dark-energy density, and $\Omega_M$ the matter-energy density, is given by:

\begin{equation}
\label{ch7:eq:distelement}
\frac{\partial L}{\partial z} = \frac{c}{H_0} \frac{1}{1+z} \frac{1}{\sqrt{\Omega_{\Lambda} + \Omega_M (1+z)^3}}  {\text  .}
\end{equation}

The Breit-Wheeler formula provides the pair creation cross section as:

\begin{equation}
\label{ch7:eq:crosssec}
\sigma_{\gamma\gamma}(\beta) = \frac{3 \sigma_T}{16}  (1-\beta^2) \left[ 2\beta (\beta^2 -2) + (3 - \beta^4) \ln \left( \frac{1 + \beta}{1 -\beta} \right) \right] {\text  ,}
\end{equation}

\noindent where $\sigma_T$ is the Thomson cross-section, and $\beta$ is given by:

\begin{equation}
\centering
\label{ch7:eq:beta}
\beta = \sqrt{1 - \frac{2(m_{\rm e} c^2)^2}{E_\gamma' \epsilon'} \frac{1}{1-\cos \theta}} {\text  ,}
\end{equation}

The attentive reader should note that $E_\gamma' = E_\gamma\times(1+z)$ and $\epsilon'$ are the energies of the gamma-ray and EBL photon interacting at redshift $z$. It is worth noting that the physical condition that $\beta$ be a real number and that $\sigma_{\gamma\gamma}\geq0$ imposes the threshold condition:

\begin{equation}
\label{ch7:eq:EBLthresholdAdvanced}
E_\gamma' \epsilon' \geq \frac{2\left(m_{\rm e} c^2\right)^2}{1-\cos \theta} {\text  ,}
\end{equation}
which translates to the lowest possible threshold for head-on collisions, i.e.\ Eq.~\eqref{ch7:eq:EBLthreshold} for $\theta=\pi$.

As illustrated in Fig.~\ref{ch7:fig:EBLspectrum}, the pair-production cross section, integrated over the line of sight,\footnote{More accurately, we refer here to the quantity called ``EBL kernel'' \citep{2015ApJ...812...60B}, properly normalized so that its unit is the inverse of that of the EBL intensity.} has a full-width at half maximum covering almost a decade in EBL photon energy and peaking at an energy about twice that obtained from the threshold condition in Eq.~\eqref{ch7:eq:EBLthreshold}. This results in gamma rays interacting preferentially with EBL photons in the wavelength range:
\begin{equation}
\label{ch7:eq:EBLwindow}
\lambda_{\rm EBL} \simeq 0.5-5\:\mu{\rm m} \times \left(\frac{E_\gamma}{1\:{\rm TeV}}\right)\times (1+z)^2
\end{equation}
where the factor $(1+z)^2$ accounts for the redshift of the energies of the photons. Interaction with photons from the CMB, which reaches an intensity comparable to that of the CIB around $\sim250\:\mu$m, becomes relevant for gamma-ray energies larger than $\sim 50$\:TeV. From Eq.~\eqref{ch7:eq:EBLwindow}, one can infer that measurements of gamma-ray absorption around 20\:TeV, obtained from nearby sources ($z\lesssim0.05$), probe the CIB up to 100\:$\mu$m, while distant sources ($z\gtrsim 1$) detected up to a few hundred GeV probe the UV-optical part of the COB. 

\begin{figure}[t]
\centering
\includegraphics[width=0.8\textwidth]{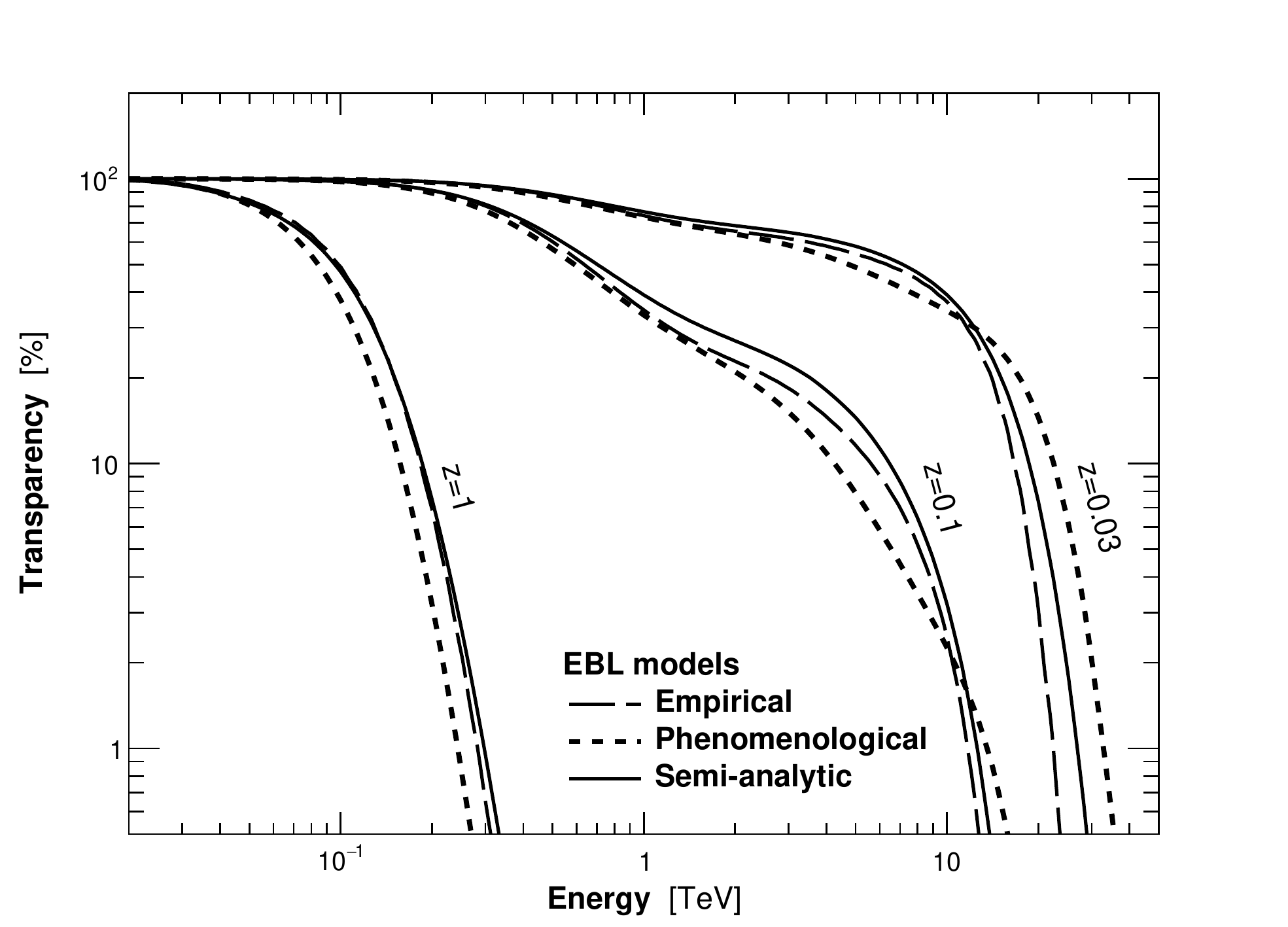}
\caption{Attenuation factor, in percent, as a function of gamma-ray energy on Earth for sources located at $z=0.03$, $z=0.1$, and $z=1.0$, following the empirical model of \citet{2011MNRAS.410.2556D}, the phenomenological model of \citet{2010ApJ...712..238F}, and the semi-analytical model of \citet{2012MNRAS.422.3189G}.}
\label{ch7:fig:EBLopticaldepth}
\end{figure}

The transparency of the universe to gamma rays scales with the attenuation factor $\exp\left(-\tau(E_\gamma,z)\right)$. At first order, the optical depth is a linear function of energy and of EBL photon density integrated over the line of sight, so that the transparency to gamma rays decreases with increasing energy and source distance, as illustrated in Fig.~\ref{ch7:fig:EBLopticaldepth}. Substructures can be identified in Fig.~\ref{ch7:fig:EBLopticaldepth}, with inflection points tracing the variations of the EBL spectrum as a function of wavelength. The features observed in the attenuation reflect the dependence of the EBL intensity on wavelength, with low-energy gamma rays (up to $\sim 1\:$TeV) interacting with photons from the COB while higher-energy gamma rays interact mostly with photons from the CIB.

The gamma-ray spectrum observed from an extragalactic source is the product of the intrinsic spectrum of the source and of the attenuation factor, $\exp\left(-\tau(E_\gamma,z)\right)$. The intrinsic spectrum is not \textit{a priori} known, thus multiple features can be used to constrain or measure the EBL density: the variation of the spectral index from the unabsorbed region to the attenuated one (the $\Delta\Gamma$ method, e.g. \citealp{2013A&A...554A..75S}), the measurement of the cosmic gamma-ray horizon (the energy for which $\tau=1$, e.g. \citealp{2013ApJ...770...77D}), and the reconstruction of the full spectral signature that is shown in Fig.~\ref{ch7:fig:EBLopticaldepth}~\citep{2012Sci...338.1190A,2013A&A...550A...4H}. The exploitation of these features and the resulting constraints are further discussed in the next section. 

\subsection{VHE constraints}

The extragalactic gamma-ray sky at GeV and TeV energies is dominated by blazars (BL~Lac objects and flat-spectrum radio quasars), which have been detected by ground-based instruments to redshifts of about $z=1$. The energy spectra of less distant sources can extend to high energies, in some cases to tens of TeV for sources with particularly hard intrinsic spectra (see \citealp{2020NatAs...4..124B} for a review). Many blazars show dramatic flux variability. For example, up to ten thousand gamma rays have been collected in less than an hour during the highest states of the blazar PKS~2155-304 \citep{2007ApJ...664L..71A}, enabling the construction of energy spectra with small statistical uncertainties. Blazars thus appear to be ideal gamma-ray beacons to probe the EBL. Long gamma-ray bursts (GRBs), a new class of VHE emitting sources, are also promising targets~\citep{MAGICGRB, 2019Natur.575..464A, 2021Sci...372.1081H}. The current generation of imaging atmospheric Cherenkov telescopes (IACTs), together with ~\LAT\ \citep{2012Sci...338.1190A}, has mostly used blazar observations to probe the EBL spectral energy distribution (SED) (see e.g. \citealp{2019ApJ...874L...7D}).

Individual measurements of sources at boundaries of the TeV-accessible redshift range ($z\sim 1$) are relevant for testing the redshift evolution of the EBL. Both MAGIC \citep{2015ApJ...815L..23A,2016A&A...595A..98A} and VERITAS~\citep{2015ApJ...815L..22A} have detected flat-spectrum radio quasars at $z>0.9$. Studies of the spectra of these sources have placed competitive constraints on the optical peak in the EBL SED. These measurements can be compared to EBL constraints obtained with nearby sources, which probe similar optical depths at the high-energy ends of their spectra. The comparison of measurements with low- and high-redshift gamma-ray sources provides an important consistency check for our understanding of the EBL evolution, provided that the EBL SED is well-constrained.

The most robust measurements use large ensembles of blazar spectra, mitigating the sensitivity to any unexpected spectral properties affecting individual sources. These include absorption of gamma rays by local radiation fields in the vicinity of the source \citep{2007ApJ...665.1023R} and unaccounted-for variations in spectral shape over time. 

While the latest EBL measurements have focused on a few well-tested methods, we pause briefly to describe alternative strategies. In one approach, intrinsic spectra in the VHE range were predicted from the blazars' multiwavelength SEDs (from radio to \LAT\ measurements), assuming that the SEDs could be described by synchrotron self-Compton models. The optical depth as a function of energy and cosmic gamma-ray horizon were calculated from the predicted and observed VHE spectra, and compared against EBL model predictions~\citep{2013ApJ...770...77D}. The observed values were found to be consistent with model predictions.

Another method utilizes features in the optical depth as a function of energy. It is visible from Fig.~\ref{ch7:fig:EBLopticaldepth} that there is a flattening in the optical depth around 1\:TeV (most visible for $z=0.1$). This flattening should propagate to a significant change (a ``break") in spectral shape around 1\:TeV for a precisely-measured energy spectrum. Testing for a spectral break in sources located at different redshifts thus provides an alternative method for probing the spectral shape of the EBL. Such a study was performed using \LAT\ and VHE spectra \citep{Orr2011}, and the authors estimated the EBL intensity at 15\:$\mu$m to be $1.36 \pm 0.58$~nW m$^{-2}$ sr$^{-1}$, while demonstrating agreement with model predictions at other wavelengths.

The study of \citet{2013A&A...550A...4H} established the first significant detection of the EBL imprint at VHE and developed the basis of a method now commonly employed by the community. More recently, H.E.S.S. made a multi-source measurement of the EBL imprint using 21 spectra from nine blazars, covering a redshift range from $z=0.031$ to $z=0.287$ and an energy range from 190\:GeV to 19.5\:TeV~\citep{HESSEBL2017}. Datasets were divided into low- and high-flux states, resulting in multiple spectra per source for strongly variable sources, an approach, now standard, to address flux-level-dependent spectral variability. The method of measuring the EBL was to fit the spectra with a log-parabolic (or power law, in the case of no curvature) shape multiplied by an EBL attenuation factor of the form $\exp\left(-\tau(E_\gamma,z, \rho_{i})\right)$. Crucially, the level of the EBL intensity, denoted by the parameter $\rho$, was allowed to vary independently in several adjoining energy bands, indicated by the index $i$. This approach does not make any assumption about the shape or normalization of the EBL. The evolution of the EBL with redshift is tuned to replicate the predictions of \citet{2008A&A...487..837F}; this injects some dependence on a theoretical EBL model, although the impact is limited for gamma-ray sources with redshifts $z<0.7$ \citep{2015ApJ...812...60B}. 

\begin{figure}[t]
\centering
 \includegraphics[width=0.8\textwidth]{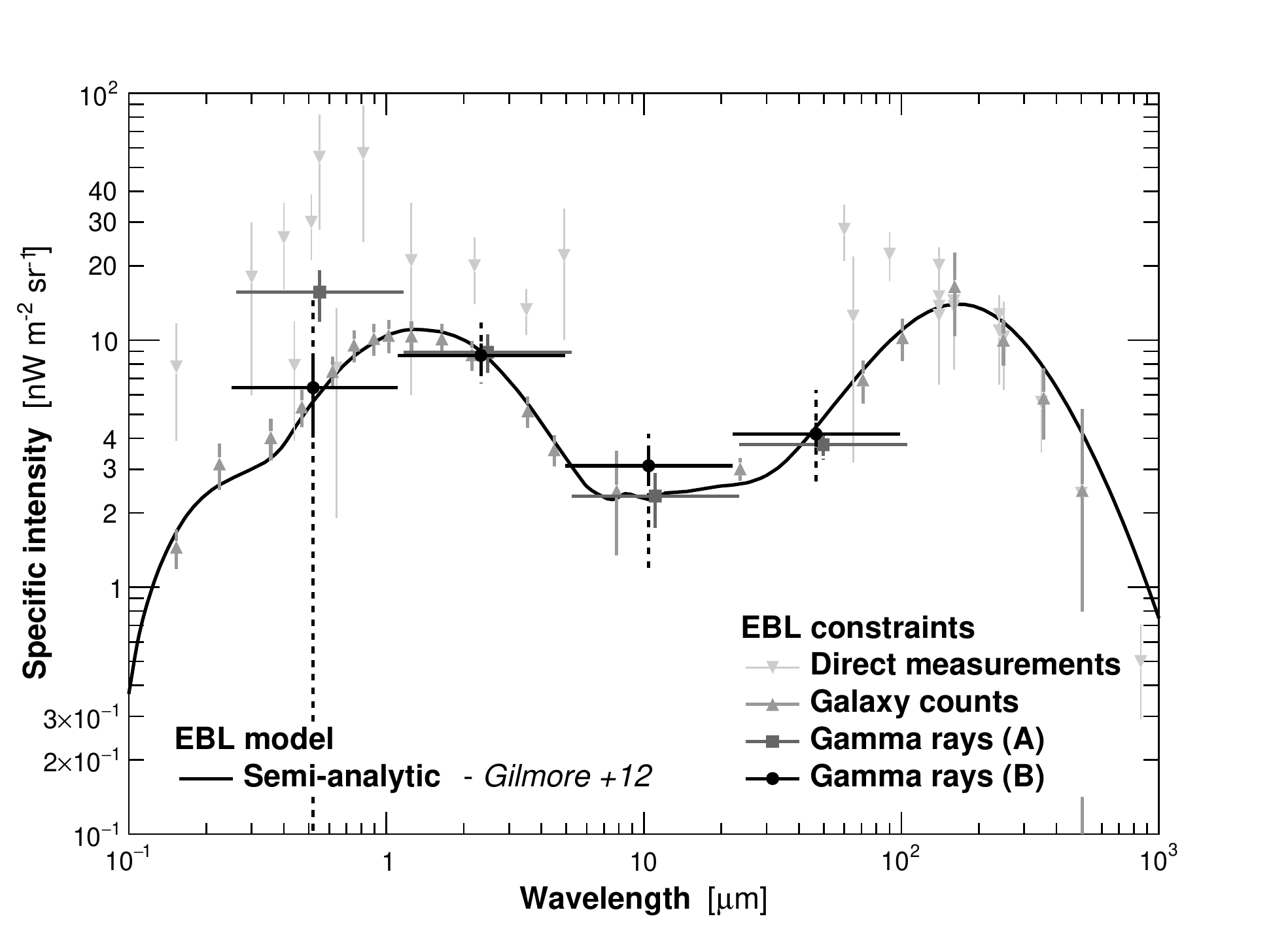}
\caption{Observational constraints on the spectrum of the EBL at $z=0$. Lower limits obtained from galaxy counts, which could miss unresolved populations or any truly diffuse EBL component, are shown as grey upward-pointing triangles \citep{2016ApJ...827..108D}. Constraints from direct measurements, possibly contaminated by foregrounds, are shown as light-grey downward-pointing triangles \citep{2013APh....43..112D}. Gamma-ray measurements are taken from the works of \citet{2015ApJ...812...60B} (A) and \citet{HESSEBL2017} (B). The semi-analytic model of \citet{2012MNRAS.422.3189G} is shown as a continuous line.}
\label{ch7:fig:EBLmeasuredspectrum}
\end{figure}

The combined best fit to all spectra reflects both the shape and normalization of the optical depth as a function of energy. The best fit including EBL attenuation was strongly favoured over a fit assuming no EBL attenuation, indicating evidence for a non-zero EBL at 9.5$\sigma$. The optical depth as a function of energy was translated back into an EBL intensity as a function of wavelength. The resulting EBL SED is shown in Fig.~\ref{ch7:fig:EBLmeasuredspectrum} (open squares labelled ``B"). Both statistical and systematic uncertainties are shown. The latter are dominated by uncertainties on the H.E.S.S. energy scale. 

Similar approaches to measuring the EBL were adopted by the MAGIC and VERITAS Collaborations \citep{2019MNRAS.486.4233A, 2019ApJ...885..150A}, using their own sets of blazar spectral measurements. In the case of MAGIC, contemporaneous \LAT\ spectra were also included, providing a more constraining fit, albeit with larger systematic uncertainties. VERITAS used a different analysis approach to that of MAGIC and H.E.S.S., testing the properties of a set of generic, smoothly varying EBL SED shapes. The results of both of these studies were consistent with the H.E.S.S. results.

A different, but also model-independent, approach was taken in a measurement of the EBL based on a large sample of archival spectral measurements \citep{2015ApJ...812...60B}. Thirty sources and 86 spectra were analysed, bringing together spectral measurements from all of the current IACTs, as well as older instruments. Additionally, information from \LAT\ spectra was used, namely the requirement that the VHE spectra be softer than the contemporaneous \LAT\ spectra. 

The EBL signature was detected at the 11$\sigma$ confidence level, and the SED resolved as shown in Fig.~\ref{ch7:fig:EBLmeasuredspectrum} (filled squares labelled ``A"). This strong detection and the relatively small errors on the points describing the EBL shape made it possible to disfavour several theoretical EBL models. The assessment of systematic uncertainties was, however, complicated by the use of archival measurements from multiple instruments. It should be noted that instrumental effects can be more thoroughly accounted for in the analysis of lower-level data.\footnote{Besides the VHE energy scale, instrumental effects relevant to gamma-ray studies of the EBL in particular include energy resolution, and selection effects related to the choice of spectral points included in the analysis.}

Selected results are summarised in Fig.~\ref{ch7:fig:EBLmeasuredspectrum}. For clarity, Fig.~\ref{ch7:fig:EBLmeasuredspectrum} only shows the semi-analytic model of \citet{2012MNRAS.422.3189G}. While no tension with the empirical model of \citet{2011MNRAS.410.2556D} was found in \citet{2015ApJ...812...60B}, the phenomenological model of \citet{2010ApJ...712..238F} appears to be disfavoured due to an underestimation of the CIB amplitude, also suggested by the integrated galaxy light measured above $100\:\mu$m \citep{2016ApJ...827..108D,2020MNRAS.491.1355D}. The most up-to-date empirical \citep{2017A&A...603A..34F, 2021MNRAS.507.5144S} and phenomenological models \citep{2018MNRAS.474..898A}, driven by the recent gamma-ray detections and updated constraints from galaxy counts \citep{2016ApJ...827..108D}, show an agreement with the semi-analytic model of \citet{2012MNRAS.422.3189G} displayed in Fig.~\ref{ch7:fig:EBLmeasuredspectrum} at the $10-20\:\%$ level in the far-infrared region. This level of agreement suggests that both the models and the galaxy counts measurements are converging to comparable intensities not only in the wavelength region of the COB but also of the CIB.

We note that the current generation of IACTs has not reached the physical limit for resolving structures in the EBL spectrum, imposed by the natural width of the EBL kernel. Both of the measurements shown in Fig.~\ref{ch7:fig:EBLmeasuredspectrum} are based on large data samples, spanning years of observations and including bright flares. In spite of this, the EBL intensity can be extracted for only a limited range of EBL wavelengths, leaving ample room for progress with future observations.

\subsection{Relation to non-VHE measurements}
Our knowledge of the EBL benefits from the existence of a variety of measurement techniques beyond the VHE approach described above. Non-VHE measurements can be divided into direct measurements and constraints from galaxy counts (see \citealp{2019ConPh..60...23M, 2021arXiv210212089D} for reviews). 

Direct measurements attempt to observe the sky (usually from space, as far from Earth as Pluto's orbit for the New Horizons mission), and subtract off foreground contamination from non-EBL light (e.g. \citealp{2017ApJ...839....7M, 2017NatCo...815003Z}). This includes a number of light sources, some of which are diffuse/isotropic: stars, scattered light in Earth's atmosphere, and light scattered by dust in the solar system (zodiacal light) and in the Milky Way. Due to the difficulty of adequately modelling and controlling against the foregrounds, direct measurements are generally treated as upper limits on the EBL intensity. Indeed, it has been pointed out that the spectrum of the EBL inferred from direct measurements shows similarities with the zodiacal light spectrum, suggesting an incomplete foreground subtraction~\citep{0004-637X-635-2-784}. An alternative approach that does not rely on the subtraction of a model-dependent foreground consists in measuring the night-sky brightness in the vicinity and within the shadow of a dark, opaque nebula. This approach has been used to measure the EBL intensity in the optical range and its results are consistent with integrated galaxy light and VHE measurements, albeit with larger uncertainties \citep{2017MNRAS.470.2152M, 2017MNRAS.470.2133M}. 

On the other hand, robust lower limits on the EBL intensity can be derived by summing the observed number of galaxies, accounting for their brightness, as a function of wavelength \citep{0004-637X-683-2-585, 2000MNRAS.312L...9M, 2016ApJ...827..108D}. These measurements have become increasingly precise as deep surveys have resolved fainter and fainter galaxies. One of the best examples to date lies in the $0.4-2\:\mu$m range, where a precision on integrated galaxy light close to 5\% has been claimed \citep{2021MNRAS.503.2033K}. However, such measurements are incapable of resolving diffuse emission. By comparing them to the VHE measurements, one can test for the presence of a diffuse component. It is interesting to note from Fig.~\ref{ch7:fig:EBLmeasuredspectrum} that the EBL intensity derived from VHE measurements does not differ significantly from the integrated galaxy light. This rules out a large diffuse component of the EBL, although the uncertainties on both sets of measurements still allow for some diffuse contribution at the level of a few nW\:m$^{-2}$\:sr$^{-1}$.

The spectrum and evolution of the EBL are tightly connected with the history of star formation, including that of core collapse supernovae. The cumulative emission of neutrinos in the MeV energy range from all these objects builds up the diffuse supernova neutrino background \citep{2009PhRvD..79h3013H}. A signal from the latter could start to emerge in future observations e.g.\ of the Super-Kamiokande Gadolinium experiment \citep{2016APh....79...49L}, providing constraints not only on the rate of core collapse supernovae but also on the CSFH and the initial mass function. The interplay of MeV neutrino measurements with high-precision EBL measurements provides exciting prospects for the understanding of the evolution of baryonic matter in the universe.

Finally, gamma-ray constraints on the EBL, particularly in the FIR band, could help further constrain the propagation of other astrophysical particles: ultra-high-energy cosmic rays (UHECR), which are nuclei with energies larger than $10^{18}$\:eV ($\equiv1$\:EeV). UHE protons interact with CMB photons through the so-called Greisen-Zatsepin-Kuzmin (GZK) effect, losing energy via pion and electron-positron pair production. Heavier nuclei, ranging from He to Fe, can also photodissociate on low-energy photon fields, i.e.\ release a neutron or a proton carrying a fraction of the energy of the parent nucleus. Photodissociation on FIR EBL photons is the dominant energy-loss process for low-to-intermediate mass nuclei with a Lorentz factor $\Gamma\sim10^9$ \citep{2015JCAP...10..063A}. The uncertainty on the intensity of the CIB, as shown in Fig.~\ref{ch7:fig:EBLspectrum}, has a non-negligible impact on the interpretation of the UHECR spectrum above a few tens of EeV. It additionally impacts the flux of cosmogenic neutrinos, i.e.\ those produced along the line of sight by UHECR, particularly around $10^{17}$\:EeV.

\subsection{Connecting the EBL and the Hubble constant}

An enticing inversion of EBL measurements is to use the same VHE gamma-ray propagation effects to measure the Hubble constant, $H_{0}$. Proposed from the early days of gamma-ray astronomy \citep{1994ApJ...423L...1S, BLANCH2005588, BLANCH2005598, BLANCH2005608, 2008MNRAS.389..919B}, this alternative method of measuring $H_{0}$ remains relevant, as tension between different measurements of $H_{0}$ persists at a significance currently estimated to be larger than $4\sigma$ \citep{2021CQGra..38o3001D}. Two approaches have been employed to constrain the universe's expansion rate with measurements of gamma-ray absorption. 

The first approach, largely independent of the EBL model, consists in comparing the integrated galaxy light to the gamma-ray optical depth inferred from VHE emitters in the local universe ($z\leq 0.2$), where evolutionary effects have a negligible impact on gamma-ray absorption. Such approaches are labeled here as ``gamma-ray/local EBL''. Since the optical depth is the product of the EBL intensity, pair-production cross-section and distance element, with the latter scaling as $H_{0}^{-1}$, the Hubble constant can be constrained by assuming that the EBL consists entirely of integrated galaxy light. The latter measurements are independent of the Hubble constant, but may underestimate the total EBL, particularly if truly diffuse components or unresolved populations of UV-optical photons are non-negligible.

The second approach exploits the latest models of the EBL which aim at reproducing the galaxy counts that result in the integrated galaxy light, in addition to the CSFH and metallicity evolution for phenomenological models. Such approaches are labeled here as ``gamma-ray/CSFH''. For a fixed astrophysical model, the additional dependence of the EBL density on the Hubble constant, scaling as $H_{0}^{3}$, provides a further handle on the universe's expansion rate. This approach is more sensitive to variations of the Hubble constant but could also be more dependent on the underlying astrophysical model.

\begin{figure}[b!]
\includegraphics[width=0.75\textwidth]{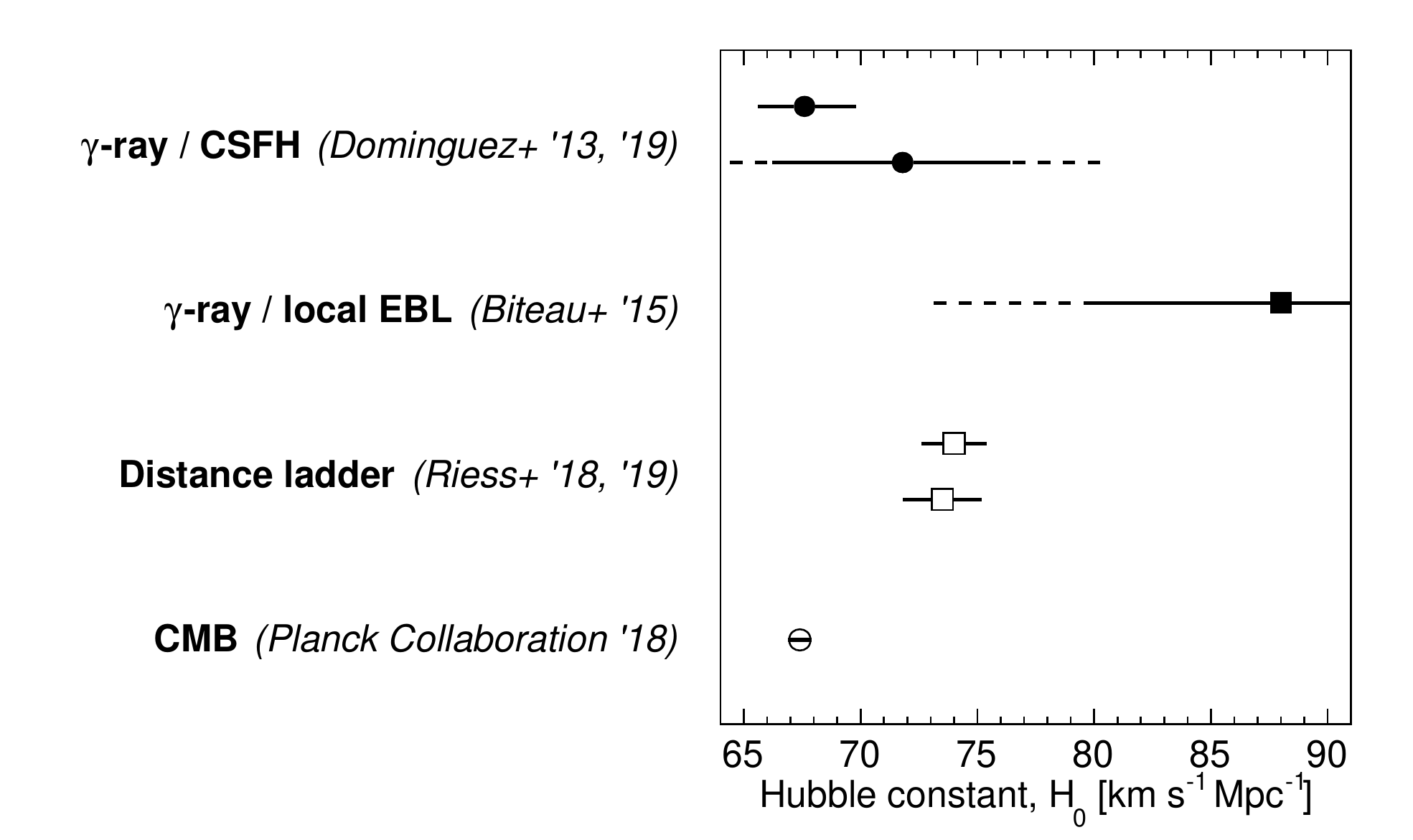}
\caption{Comparison of measurements of the Hubble constant by Planck \citep{Planck2015Hubble} and the Hubble Space Telescope \citep{HubbleHST2018} with VHE gamma-ray measurements \citep{2013ApJ...771L..34D, 2015ApJ...812...60B, 2019ApJ...885..137D}.}
\label{ch7:fig:Hubbleconstant}
\end{figure}

The first measurements exploiting the local \citep{2015ApJ...812...60B} and evolutionary techniques \citep{2013ApJ...771L..34D} were dominated by systematic uncertainties on the integrated galaxy light and EBL model, respectively, as well as those on the gamma-ray energy scale. As shown in Fig.~\ref{ch7:fig:Hubbleconstant}, the systematic uncertainties on $H_0$ are on the order of $10-15$\:km\:s$^{-1}$\:Mpc$^{-1}$, preventing a useful contribution to the debate between early-type and late-type measurements of $H_0$, which differ by about $5$\:km\:s$^{-1}$\:Mpc$^{-1}$. Recently, \citet{2019ApJ...885..137D} further developed the evolutionary approach and tested both the phenomenological model of \citet{2010ApJ...712..238F} and the empirical one of \citet{2011MNRAS.410.2556D} against  the cosmic gamma-ray horizon inferred from VHE archival data and \textit{Fermi}-LAT data. The authors claimed a measurement with an accuracy of 3\:km\:s$^{-1}$\:Mpc$^{-1}$ when fixing $\Omega_{M}$ to its reference value and with an accuracy of 6\:km\:s$^{-1}$\:Mpc$^{-1}$ when leaving $\Omega_{M}$ free (the former is reported in Fig.~\ref{ch7:fig:Hubbleconstant}). 

Independent analyses are critically needed to assess and compare the potential of such gamma-ray constraints. If confirmed, current and future gamma-ray observations could provide relevant inputs to the debate on the current cosmological paradigm. The number of extragalactic gamma-ray sources detectable by the Cherenkov Telescope Array (CTA) and the ability to measure their redshifts will be key in the development of this scientific avenue.

\subsection{Outlook}

The EBL encodes information about the CSFH and accretion of matter over cosmic ages. Ground-based gamma-ray astronomy is now mature enough to detect the imprint of the EBL in the spectra of gamma-ray blazars and measure the EBL spectrum at $z=0$ from optical to MIR wavelengths.

Nonetheless, wide areas of research on the EBL remain open. Tighter constraints on the UV part of the EBL spectrum, probed by distant sources beyond $z=1$, will provide insight into the role of the AGN in ionizing the universe \citep{2018MNRAS.474..898A}. The signature of polycyclic aromatic and amorphous hydrocarbons, responsible for the bump around 15\:$\mu$m in the models shown in Fig.~\ref{ch7:fig:EBLspectrum}, remains undetected at multi-TeV energies. Theoretical and observational  uncertainties in the FIR band, which could be solved by high-quality measurements beyond 10\:TeV, still hinder a full grasp of UHECR propagation. Exciting prospects can also be found in the detection of the spectral signature of the first stars, microquasars or outflows that may have been responsible for the reionization of the universe. The signature is expected to be visible as a broad feature above 1\:$\mu$m, in the core EBL wavelength range probed by TeV instruments~\citep{2012MNRAS.420..800G}. Finally, the wavelength range close to the peak of the COB offers interesting prospects for constraining the amount of light emitted at the outskirts or even outside of galaxies, by stars in the intra-halo, intra-group and intra-cluster media \citep{2019ConPh..60...23M}.

Besides providing a wide wavelength coverage of the EBL thanks to its order of magnitude improvement in sensitivity with respect to current-generation instruments, CTA will also enable the study of the evolution of the EBL, by reconstructing the absorption in successive layers of redshift \citep{2021JCAP...02..048A}. The expected accuracy on the EBL reconstruction is two-to-three times better for CTA than for current-generation instruments, and it is anticipated that EBL constraints will be derived with gamma-ray sources located at redshifts as distant as $z=2$. This will enable ground-based gamma-ray astronomy to place meaningful constraints on the CSFH, as recently done by \citet{2018Sci...362.1031F}. The low-energy threshold of CTA will further help in disentangling intrinsic spectral curvature due to internal absorption processes, energy cut-offs related to particle acceleration and escape, and extrinsic spectral features due to the propagation of gamma rays on cosmological scales. Finally, it is expected that CTA will set robust constraints on the Hubble constant, testing the current cosmological paradigm. 

\section{The intergalactic magnetic field}
Understanding the generation and evolution of large-scale magnetic fields in the universe is a key topic in cosmology, and one for which persistent questions remain. Magnetic fields permeate the universe from scales as small as stars, to galaxies and galaxy clusters, to the filaments separating the voids in the large-scale structure, and at the largest scales, to the voids themselves. 

Two possibilities have been proposed for the origin of cosmological magnetic fields: primordial and astrophysical origin. In the former case, a weak magnetic field is generated in phase transitions in the early universe. This ``seed'' field is then amplified in astrophysical systems to the strong magnetic fields (${\sim}\:\mu$G or greater) observed in galaxies and galaxy clusters. In the latter case, the magnetic fields are generated and amplified within galaxies and then ejected beyond the immediate environment of the astrophysical accelerator into the intergalactic medium. The timing of the magnetic-field generation is notably different in the two cases: shortly after the Big Bang, or after the onset of galaxy formation. This difference propagates to differences in the magnetic-field strength, correlation length (distance over which the field behaves coherently), and the evolution of the magnetic energy density with redshift in the different generation scenarios.

It is worth briefly discussing the underlying mechanisms for the generation and amplification of magnetic fields in plasmas. A magnetic field in a plasma can be spontaneously generated via the Biermann battery process. In a plasma with a temperature and density gradient, the lighter free electrons drift down the pressure gradient faster than protons. For misaligned temperature and density gradients, this creates an electromotive force that generates a magnetic field. Once created, a weak magnetic field can be amplified via dynamo mechanics. The movement and rotation of plasma in an astrophysical system stretches and compresses magnetic-field lines, amplifying the field as kinetic energy is converted into magnetic energy.

Extracting clues about magnetic-field generation from observations of magnetic fields in galaxies and filaments is extremely difficult due the uncertainties associated with the dynamics of the amplification process. On the other hand, the magnetic field in the voids of the large-scale structure---the intergalactic magnetic field (IGMF)---reflects the conditions of its generation for primordial production scenarios. In the case that the magnetic fields in the voids are not primordial, but due to plasma outflows from astrophysical systems, they would provide information about the efficiency of magnetic energy transfer into the voids. Consequently, regardless of the generation scenario, magnetic fields in voids are of particular interest \citep{Neronov2009,2013A&ARv..21...62D}.

\begin{figure}[t]
\centering
\includegraphics[width=0.75\textwidth]{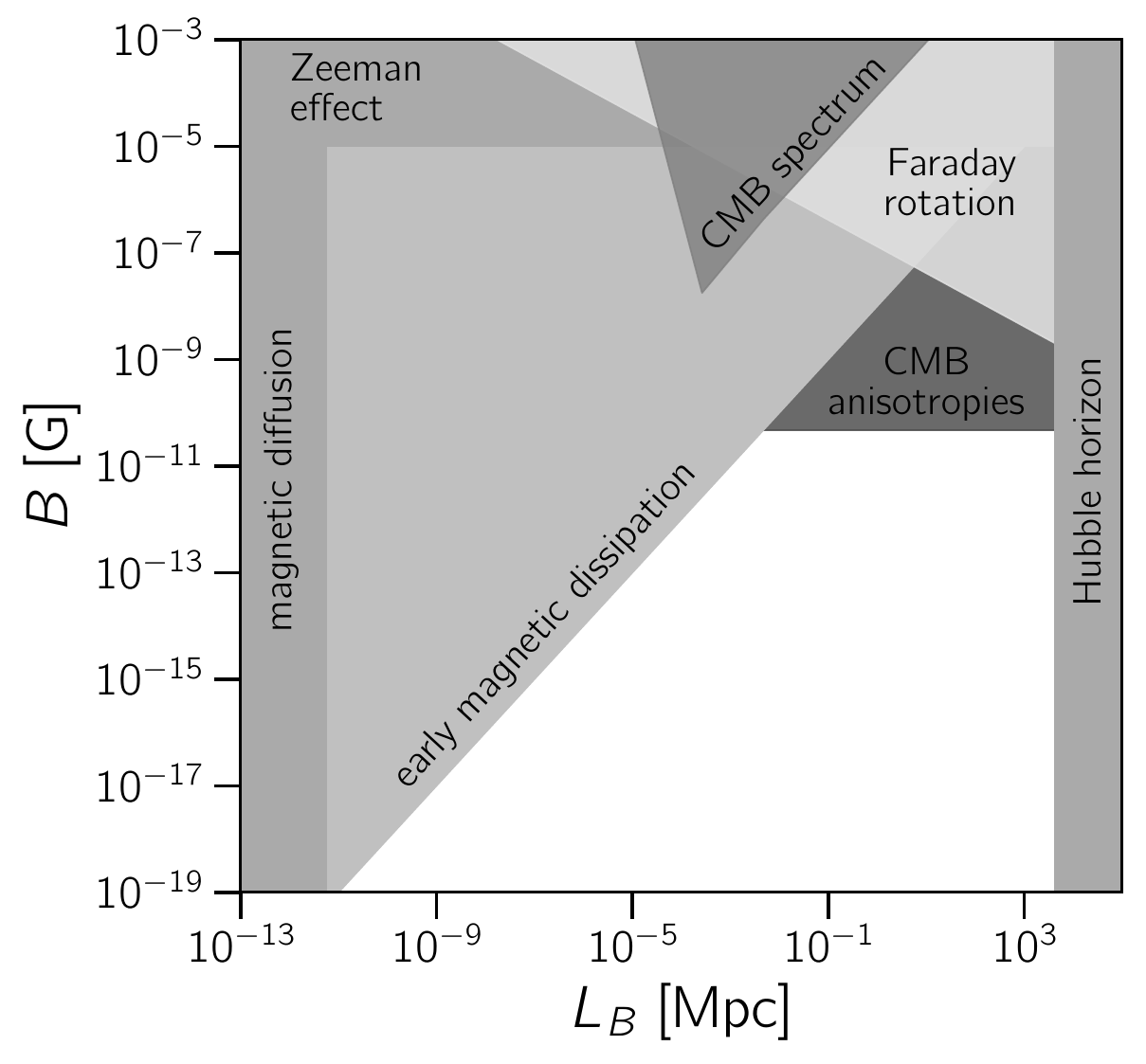}
\caption{The allowed region of IGMF strength and correlation length. Bounds from  non-VHE measurements (discussed in the text) are shown by the shaded regions. Credit: Rafael Alves Batista, adapted from \citet{2021Univ....7..223A}.}
\label{ch7:fig:IGMFallowed}
\end{figure}

Measuring the IGMF poses experimental challenges, primarily because it is expected to be extremely weak. While some experimental and theoretical bounds exist, the allowed range of strength and correlation length is distressingly broad, as shown in Fig.~\ref{ch7:fig:IGMFallowed}~\citep{2021Univ....7..223A}. The lowest guaranteed theoretical bound on the IGMF exploits the Harrison effect \citep{Harrison1970}. Magnetic fields are generated during the radiation-dominated era due to the difference in angular momentum of electron-photon and ion gases as rotating protogalaxies expand. The minimum magnetic field strength generated by this process gives a lower bound on the IGMF strength in voids of $10^{-29} \:$G for a correlation length of $10\:$Mpc. The gamma-ray measurements described below set more aggressive lower limits on the IGMF strength.

\subsection{Field generation}
\subsubsection{Early universe}
Magnetic fields can be generated in the early universe due the charge separation that occurs during phase transitions. 
The Hubble horizon at the time of field generation determines the possible range of correlation lengths for the case of magnetic fields generated after inflation. An IGMF generated during a QCD phase transition, for example, is expected to have a longer correlation length than one generated during an electroweak phase transition, although it must be noted that there is a large overlap region in the strength and correlation length predictions in the two generation scenarios (see e.g. \citealp{Neronov2009}). The correlation length of inflation-generated magnetic fields, on the other hand, is not limited by the Hubble horizon, and can reach the current horizon scale~\citep{Kahniashvili2016}.

\subsubsection{Astrophysical}
Turbulent plasma movement occurs in a number of astrophysical systems and could lead to the injection of magnetic fields into the intergalactic medium. Star formation activity and supernovae in starburst galaxies are candidates for IGMF generation. Starburst activity of early-forming dwarf galaxies at a time when the universe was smaller and more densely packed could also plausibly magnetise the intergalactic medium. Alternatively, at later stages in the universe's evolution, the giant lobes of radio galaxies and the jets of active galactic nuclei could expel magnetised plasma; indeed, much more energy is available from these sources than from the outflows of starburst galaxies. 

The expected correlation length for an astrophysically generated IGMF is shorter than for a primordial field, due to the later generation. The field strength is expected to be high in the vicinity of the generating sources, but to decay steeply toward zero in voids. The predicted strength in the voids is of order 10$^{-10}$~G or less, several orders of magnitude less than the strength in sheets and filaments~\citep{Neronov2009, 2017CQGra..34w4001V}.

\subsubsection{Beyond strength and length}
In addition to measuring the strength and correlation length of the IGMF, constraining its helicity is of particular interest, both for understanding the conditions at generation, and for modelling its evolution~\citep{Vachaspati2021}. The diffusion of magnetic energy from small scales to large scales depends on the helicity of the field, with a slower dissipation time for helical fields~\citep{Neronov2009}.

Lastly, measuring the redshift evolution of the field strength and energy density is important for tracing the generation of the field. An astrophysically generated IGMF is expected to have a lower energy density at high redshifts than in the local universe. 

\subsection{VHE observables}

We have already discussed the impact of the EBL on VHE photons travelling from distant blazars. Neglected in that discussion was the further trajectory of the electron-positron pairs produced in interactions with EBL photons. As charged particles, they are deflected by the IGMF, prior to interacting with the CMB. The electrons and positrons upscatter the CMB photons to higher energies via inverse Compton scattering. For sufficiently high energy particles, the process repeats and an electromagnetic cascade develops. A useful estimation of the energies involved is given in \citet{Neronov2009}:

\begin{equation}
E_{\gamma} = \frac{4}{3}(1 + z_{\gamma\gamma})^{-1}\epsilon_{\rm cmb}\frac{E_{\rm e}^{2}}{m_{\rm e}^{2}c^{4}} \simeq 0.32\:\textrm{TeV}~\Bigg[\frac{E_{\gamma_{0}}}{20\:\textrm{TeV}}\Bigg]^{2}
\end{equation}
for which $E_{\gamma}$ is the energy of the observed photon on Earth, $E_{\gamma_{0}}$ is the energy of the primary photon at the source, $m_{\rm e}$ and $E_{\rm e}$ are the mass and energy at pair production of the electron or positron, and $\epsilon_{\rm cmb}=0.6\:\textrm{meV}(1+z_{\gamma\gamma})$ is the CMB photon energy. A $\sim\:$10\:TeV primary photon is thus reprocessed to $\sim\:$100\:GeV, not far above the energy threshold of current-generation IACTs.

Deflection of the electrons and positrons by the IGMF broadens the cascade in space and delays the cascade arrival time at the observer by increasing the path length from the pair-production site to the observer. Both of these effects make it possible to probe the IGMF strength along the line of sight with observations of the VHE gamma-ray emission of distant blazars. 

The narrowly beamed ensemble of relativistic electron/positron pairs constitutes a plasma, and can develop electromagnetic and electrostatic plasma beam instabilities. Whether these instabilities develop faster than the timescale for inverse Compton scattering, and how efficiently they dissipate energy into the intergalactic medium, is important for understanding whether inverse Compton scattering is the dominant cooling process and whether an electromagnetic cascade can indeed develop \citep{BroderickPlasmaInstability}. Some simulations indicate that beam cooling from plasma instabilities is not expected to be a large effect in the case of cascades from TeV blazars \citep{Sironi2014,Rafighi2017plasmainstability}. However, other results indicate that the beam cooling from plasma instabilities could proceed on a similar timescale to inverse Compton cooling, reducing the expected cascade emission \citep{Vafin2018}.  The theoretical debate on this topic thus awaits conclusive answers. The reader should note that the results that follow assume that inverse Compton scattering is the dominant cooling process.

\subsubsection{Spectral signature}
Due to the reprocessing of emission from higher to lower energy, the shape of the GeV-TeV spectrum is sensitive to the IGMF strength. Some fraction of electron/positron pairs are deflected out of the observer's field of view for a non-zero IGMF, reducing the number of observed GeV gamma rays and the relative ratio of GeV to TeV emission. By comparing the observed source spectrum with predicted spectra with and without considering an IGMF, it is possible to set lower limits on the IGMF strength. However, it is necessary to make some assumptions about the shape of the intrinsic source spectrum.

Blazar flux variability is an important confounding factor in such studies. Unless the GeV and TeV observations are made simultaneously, it is impossible to differentiate an IGMF-induced effect from sampling different flux states with the GeV and TeV observations.  

\subsubsection{Pair halos and magnetically broadened cascades}
A second approach for measuring the IGMF is to search for a low-energy angular extension produced by angular broadening of the cascade emission. As distant sources, even the closest blazars are expected to be point-like within the angular resolution of the current instruments. However, depending on the strength of the IGMF, two types of angular extension can occur, referred to here as pair halo and magnetically broadened cascade. The key point is that some fraction of electrons (positrons) isotropize in the magnetic field and gyrate around their point of production. The stronger the IGMF, the higher the energy of the charged particles that are isotropized. 

A pair halo refers to the case where a strong IGMF (10$^{-12}$--10$^{-9}$\:G) isotropizes a large fraction of high-energy charged particles in the cascade, causing them to accumulate around the source. The reprocessed emission arriving at the observer is consequently extremely time-delayed, on the order of hundreds of years \citep{Eungwanichayapant2009}. 

In the case of magnetically broadened cascades, a weaker IGMF (10$^{-16}$--10$^{-12}$\:G, still strong enough to produce an extension beyond the point spread function of current instruments) isotropizes only lower energy charged particles, while much of the cascade travels to the observer with time delays of the order of years with respect to the primary emission. 

Predictions for angular extensions can be arrived at through analytic approximations \citep{Neronov2009, Dermer2011}. An approximation of the deflection angle of an inverse Compton gamma ray illustrates the important factors \citep{Neronov2007}: 

\begin{equation}
\theta_{ext} \simeq \frac{0.7^{\circ}}{\tau(E_{\gamma_{0}},z)}\Bigg[\frac{E_{\gamma}}{6~\textrm{TeV}}\Bigg]^{-1}\frac{B}{10^{-13}~\textrm{G}}  {\text  .}
\end{equation}
As expected, the extension scales with the IGMF strength \textit{B}, and inversely with the energy of the upscattered photon $E_{\gamma}$. It also depends on the optical depth for the primary photon (the equation holds for $\tau>1$). Consequently, the more distant the source and higher the energy of the primary photon, the smaller the predicted extension. This only holds to a point: if the mean free path of the primary gamma ray has the same scale as the distance to the observer, the cascade will not have developed enough to be observable by the time it reaches the observer. Consequently, sources at $z\simeq$ 0.1--0.2 are particularly suitable for IGMF studies in the energy range of interest for IACTs. 

\subsection{Simulating cascades}
In addition to analytical approximations, there are an abundance of numerical simulations, which treat the cascade process and magnetic-field structure with varying levels of detail \citep{Eungwanichayapant2009,Taylor2011,KACHELRIE20121036,Arlen2014,HESSExtendedEmission,Finke2015,1475-7516-2016-05-038,VERITASExtendedEmission,doi:10.1093/mnras/stw3365}. Numerical simulations, although often computationally intensive, provide more accurate predictions of the cascade spectrum and angular extension, and the ratio of cascade to total emission (cascade fraction). They also allow a detailed treatment of the primary spectrum of the source, most commonly taken as a power law with an exponential cutoff at high energy in the literature. The impact on the cascade emission of factors such as the Doppler factor of the blazar jet, the viewing angle of observer with respect to the jet, and the choice of EBL model can also be studied. 
 
\subsection{VHE constraints}
Similarly to EBL studies, hard-spectrum blazars, particularly extreme TeV blazars, are good candidates for IGMF studies \citep{2020NatAs...4..124B}. As discussed previously, the primary emission is reprocessed to much lower energies through the cascade development. The cascade emission coming from sources with soft spectral indices and low-energy (sub-TeV) spectral cutoffs falls below the GeV-TeV energy range. Extreme blazars thus represent the most viable candidates for setting VHE constraints on the IGMF.

The constraints presented below can be best read as order-of-magnitude estimates. The uncertainties on predictions of the cascade emission are still large and have a strong impact on the derived constraints. 

\subsubsection{Constraints from spectral measurements}
Many attempts have been made to set lower limits on the IGMF strength based on the GeV-TeV broadband spectrum. We only discuss selected results here. All studies nominally assumed a correlation length $\lambda_{B}\geq1\:$Mpc.

\begin{figure}[t]
\centering
\includegraphics[width=1.0\textwidth]{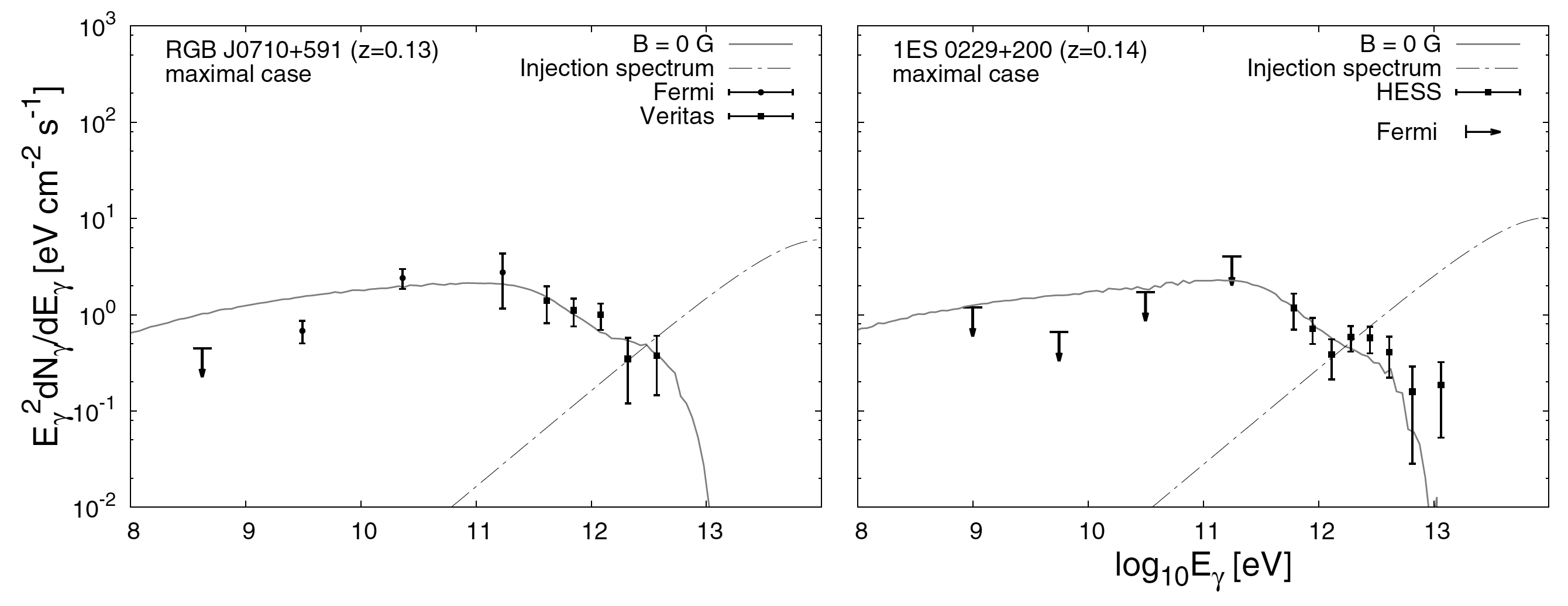}
\caption{The ~\LAT/VHE spectra for two gamma-ray blazars, shown with the predicted cascade component in the absence of IGMF (solid line). The predicted GeV component of the spectrum is higher than the observed flux. The upper bounds denoted by the arrows show the 99\% confidence level. The dashed line denotes the injection spectrum, prior to the cascade development. Figure extracted from \citet{Taylor2011}.}
\label{ch7:fig:IGMFspectralcomparison}
\end{figure}

An early study used archival TeV spectra of three hard-spectrum blazars together with simultaneous \LAT\ observations to produce broadband spectra~\citep{Taylor2011}. These were then fit with numerical predictions for the spectral shape with and without IGMF. Example spectra are shown for illustration in Fig.~\ref{ch7:fig:IGMFspectralcomparison}. As shown in the figure, the predicted spectrum for IGMF strength $B =0$ overshoots the observed \LAT\ flux. Assuming the GeV deficit is due to the cascade emission being deflected out of field of view of the observer, $B$ must be at least ${\sim}\:10^{-15}\:$G. Taking the pessimistic assumption that the source was only active during the observation time (three years at the time the study was made) and the GeV deficit is due to the time delay of the cascade emission with respect to the primary emission, a lower limit of $B\gtrsim 10^{-17}$\:G was obtained.

A later study used the four blazar datasets of \citet{Taylor2011} as well as four additional hard-spectrum blazars~\citep{Arlen2014}. Here, a scan was made of the possible parameters (jet opening angle, Doppler factor, etc.) within reasonable limits. Accounting for these uncertainties, the authors concluded that the broadband spectra are compatible with $B=0$. Some tension was observed in the case of one source (1ES~0229+200), however it was noted that this could be resolved by shifting the assumed EBL shape within experimental uncertainties. 

An overlapping set of sources, covering observations of five hard-spectrum blazars, and including several more years of \LAT\ data than previous studies, has also been considered \citep{Finke2015}. While an analytical approach to the cascade predictions was taken, the predicted cascade spectra were checked against simulation. An IGMF strength less than $10^{-19}$\:G was ruled out at ${\geq}\:5\sigma$ confidence.

\subsubsection{Constraints from searches for angular extension}
All of the currently operating IACT collaborations have made searches for angular emission around blazars. Searches for extended emission have the advantage that they can be carried out by a single instrument. The sensitivity to extended cascade emission is driven by two factors: the energy threshold and the angular resolution of current-generation IACTs. The former determines the total measurable flux of cascade emission, the latter is directly correlated to the smallest detectable extension and the weakest detectable IGMF. A standard approach has been to compare the angular profile observed in data to the prediction for a simulated point source, where the extension is entirely due to the instrument point spread function. 

An early attempt was made by MAGIC to constrain the IGMF with a search for extended emission around the nearby blazars Markarian~501 and Markarian~421 \citep{2010A&A...524A..77A}. In lieu of the detection of an IGMF signature, placing constraints is complicated by the close proximity of the sources, their extreme variability, and the presence of source-intrinsic spectral cutoffs. 

\begin{figure}[htp!]
\centering
\includegraphics[width=0.7\textwidth]{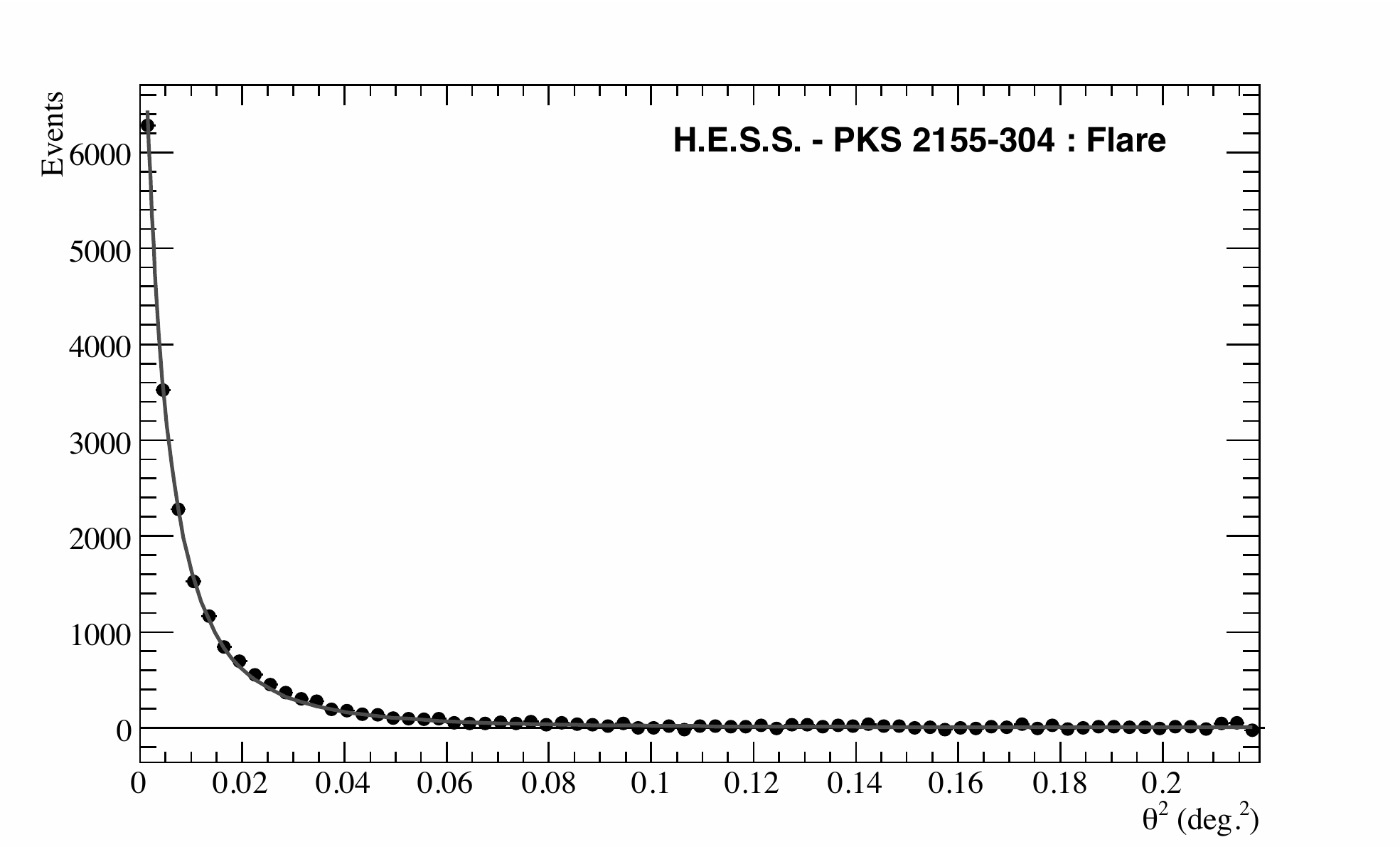}\\
\includegraphics[width=0.7\textwidth]{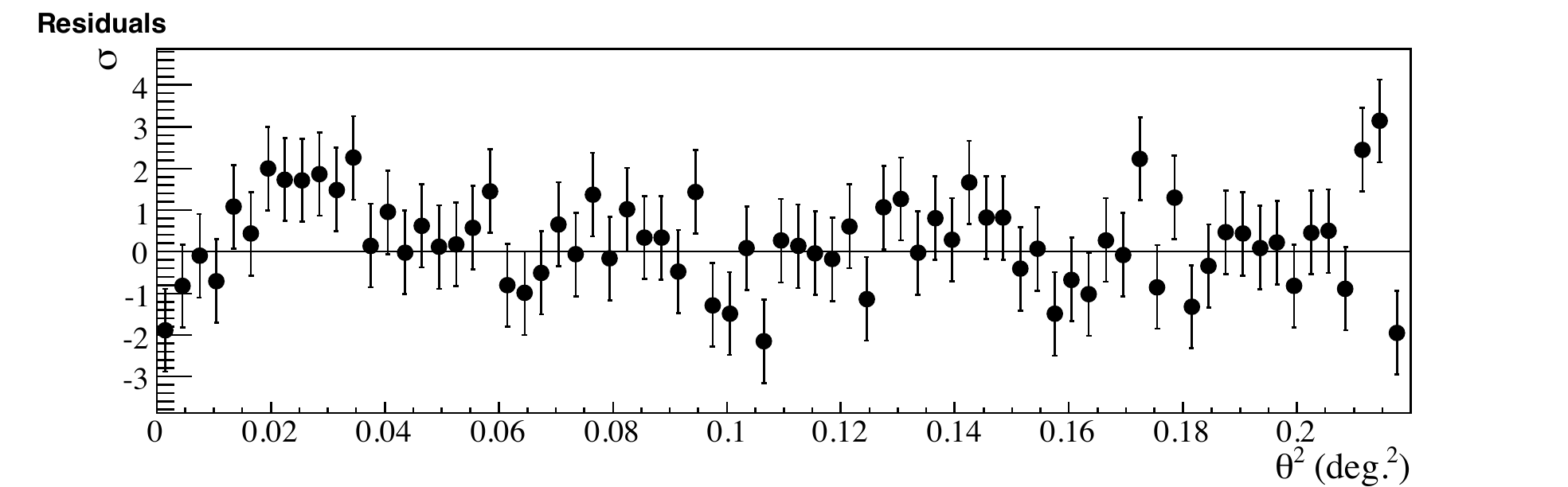}\\
\vspace{-1cm}
\hspace{-1cm}\includegraphics[width=0.59\textwidth, angle =270]{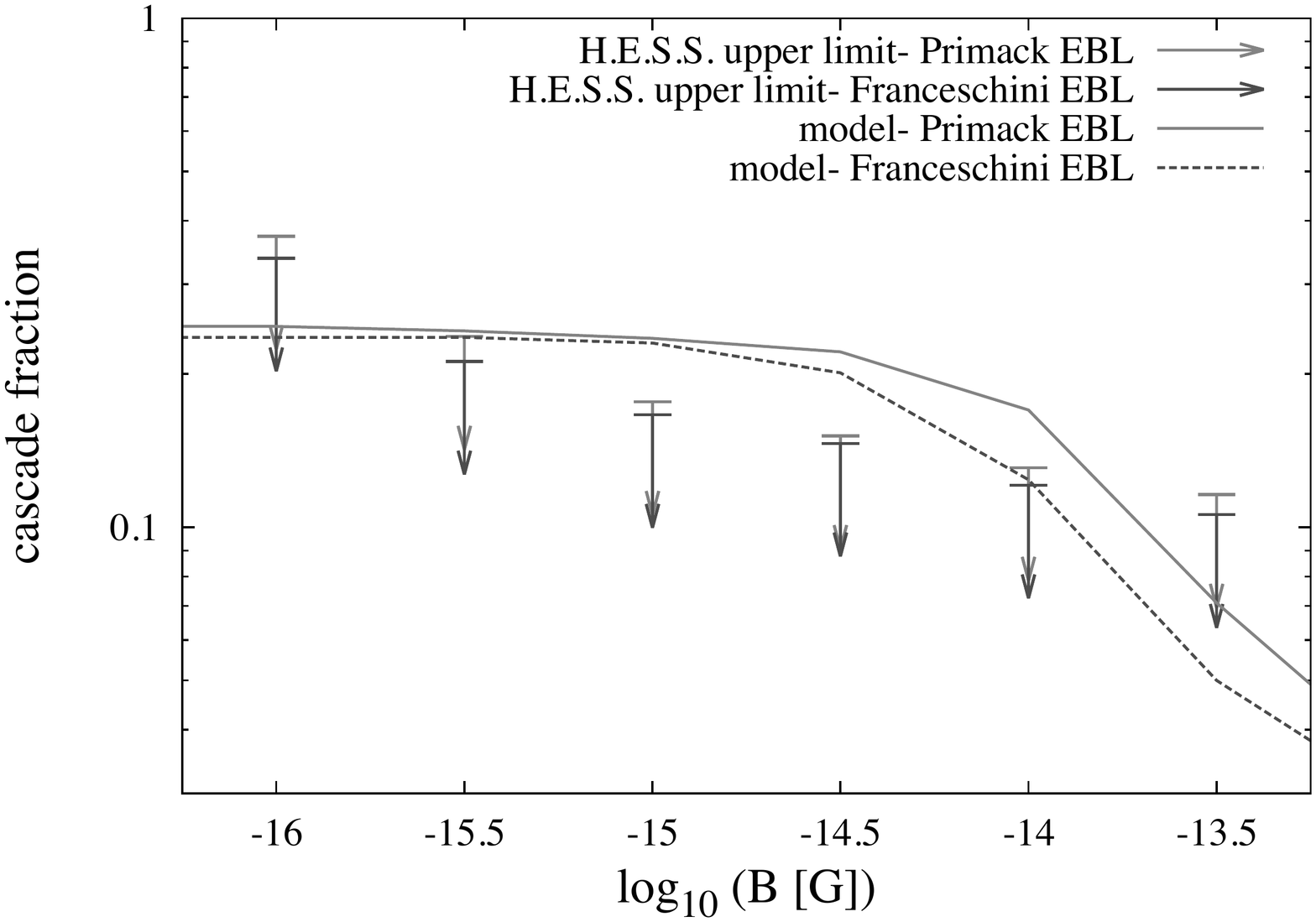}
\vspace{-1cm}
\caption{The top panel shows the angular profile of a flaring blazar observed by~\hess\ (shown as data point), whose angular extension is expected to be only due to the instrument point spread function. It is in good agreement with a simulated point source (solid line). In the bottom panel, upper limits on the observed cascade fraction are compared against predictions, as a function of IGMF strength. Figure extracted from \citet{HESSExtendedEmission}.}
\label{ch7:fig:IGMFangularextension}
\end{figure}

\hess~studied three hard-spectrum blazars located at redshifts $z=0.140-0.186$, using numerical simulations to predict the cascade flux and angular profile \citep{HESSExtendedEmission}. An example of the angular profile in data and simulation is shown in Fig.~\ref{ch7:fig:IGMFangularextension}. Extended emission was ruled out at the level of a few percent of the Crab nebula flux for the most constraining source. Additionally, \hess~set limits on the cascade fraction, which were translated into a limit on the IGMF strength. The cascade simulation predicts the cascade fraction and angular extension as a function of IGMF strength for an assumed intrinsic spectrum. Fig.~\ref{ch7:fig:IGMFangularextension} shows the evolution of the predicted cascade fraction as a function of IGMF strength, compared to upper limits on the cascade fraction extracted from data. The latter are produced by finding the maximum cascade fraction that could result in a non-detection of an angular extension. Uncertainties due to assumptions about the intrinsic spectra of the sources and the EBL model used in the cascade simulation were considered. The impact of these uncertainties is significant, but it was still possible to rule out a range of IGMF strengths around $10^{-16}$\:G to $10^{-15}$\:G at the 99\% confidence level, assuming a correlation length of 1\:Mpc.

VERITAS produced a complementary result to the \hess~measurement, and arrived at similar conclusions \citep{VERITASExtendedEmission}. Observations of seven hard-spectrum blazars were utilized, covering a range of redshifts from $z=0.031$ to $z\sim0.5$. As with the \hess~measurement, the angularly extended flux was constrained to be less than a few percent of the Crab nebula flux. One source (1ES~1218+304) was found to be the most useful object for IGMF studies, due to its brightness, redshift, and hard intrinsic spectrum with no identified cutoff. It was thus the only source considered for extracting IGMF limits. Limits on the cascade fraction as a function of IGMF strength were set, and used to constrain the IGMF strength. The choice of sources makes the \hess~and VERITAS measurements overlapping but complementary, with VERITAS ruling out a set of slightly stronger IGMF strengths between $10^{-15}$\:G and $10^{-14}$\:G at the 95\% confidence level, assuming a correlation length of 1\:Mpc.

The definitive study, at the time of writing, is a search that combined angular and spectral information from \textit{Fermi}-LAT data and archival IACT spectra~\citep{Wood2018}, with the spectral information proving key to the extraction of IGMF constraints. A set of six hard-spectrum blazars, with low detected variability and spectral points extending to an optical depth $\tau=2$ were studied individually, as well as jointly in a stacked analysis. The time period for which the sources were assumed to be active was varied from 10 to $10^{7}$ years, the former corresponding to the observation period and the latter to the typical AGN activity timescale, and the IGMF constraints re-evaluated. Even in the case of the most conservative assumption of the blazars only being active for 10 years, the extracted limits were significantly more constraining than the other measurements discussed here. A lower limit of $B\gtrsim10^{-16}$\:G was set at the 95\% confidence level, assuming a correlation length greater than 10 kpc. Combined with the smaller H.E.S.S. and VERITAS exclusion regions, the \textit{Fermi}-LAT results substantially reduce the allowed region for a weak IGMF.

\subsection{Relation to non-VHE measurements}

As is shown in Fig.~\ref{ch7:fig:IGMFallowed}, upper limits on the IGMF strength are also available based on non-VHE measurements \citep{0067-0049-151-2-271}. Magnetic fields encountered en route to the observer lead to Zeeman splitting of the 21 cm hydrogen absorption line in quasar spectra. The primary expected contributor to this is the Galactic magnetic field, thus the IGMF must clearly be less than the total magnetic field derived from Zeeman splitting \citep{Neronov2009}. Another constraint comes from polarized emission from quasars, which undergoes Faraday rotation. The polarization angle of the emission changes with wavelength, with the ratio of the change in polarization to the change in wavelength varying directly with the magnetic-field strength along the line of sight \citep{1538-4357-514-2-L79}. Lastly, tangled primordial magnetic fields are expected to produce anisotropies in the CMB, the non-observation of which can be used to constrain the field strength \citep{PhysRevD.92.123509,PlanckPrimordialFields, PhysRevLett.123.021301}. Further constraints can be derived from Big Bang nucleosynthesis, namely that a large primordial magnetic field would drive a too-fast expansion of the universe, slowing heavy-element formation and leading to a greater fraction of hydrogen to heavy elements than is observed \citep{GRASSO2001163}. 

Upper limits on the IGMF correlation length are set under the assumption that the correlation length must be less than the size of the universe, while lower limits on the correlation length can be set based on the predicted decay time of the IGMF due to magnetic diffusion.

In addition to its interest as a cosmological observable, the IGMF strength and structure is experimentally relevant for a number of measurements. Ultra-high-energy cosmic rays are bent by the IGMF, although the effect of magnetic fields in voids is expected to be subdominant compared to deflections by the stronger magnetic fields in filaments, galaxy clusters and in the Milky Way. As discussed in the chapter dedicated to fundamental physics, the IGMF is also important for studies of axion-like particles, hypothetical particles which couple to the magnetic field, converting to and from VHE photons~\citep{chapter8}.  

\subsection{Outlook}

Magnetism on cosmological scales remains a mystery. As discussed in the introduction, lower limits on the IGMF strength based on the Harrison effect are nearly fifteen orders of magnitude too low to explain the missing cascade component at low gamma-ray energies. Is the IGMF strength larger than constrained by H.E.S.S. and VERITAS? Which magnetic-field generation scenarios are the most viable and what would be the resulting strength, correlation length, and helicity of the IGMF in voids today? Could plasma instabilities play a significant role in cooling electron-positron beams that develop over dozens of Mpc?

More precise non-VHE measurements and convincing theoretical arguments may reduce the allowed region for the IGMF strength and correlation length, and further study of collective plasma effects may clarify predictions for the cascade development. However, observational searches with IACTs have already shown the possibility to probe an uncharted territory in the parameter space of the IGMF. These searches can be expected to expand beyond blazar-based studies in the coming years, as a catalog of TeV-detected GRBs is assembled, enabling a parallel set of searches. Associations between neutrinos and blazars, discussed in the chapter dedicated to multi-messenger astronomy, also present a potentially promising route to constrain the IGMF, using the time delay between the neutrino measurement and gamma-ray flare detection.

Regardless of the source class considered in IGMF searches, a low energy threshold, sensitivity from tenths to tens of TeV to constrain both the primary and secondary components, and an angular resolution at the few arcminute level constitute the key characteristics of an instrument that will be able to either measure for the first time or strongly constrain the large-scale magnetism of the universe. With these capabilities, CTA will probe IGMF strengths up to 0.3\:pG \citep{2021JCAP...02..048A}. Future VHE gamma-ray observations will doubtless improve our understanding of the origin of cosmic magnetism.

\section{Lorentz invariance violation}
We now turn to a topic focussed on the fundamental physical laws of the universe rather than its evolution. Lorentz invariance and the constancy of the speed of light $c$ for all inertial observers underpin the current understanding of relativity and quantum mechanics. A number of theories that attempt to unite gravity with the other fundamental forces and incorporate it into a quantum mechanical framework predict violation of Lorentz invariance at or below the Planck scale, $E_{\rm Pl}\approx 1.22 \times 10^{19}$~GeV \citep[e.g.][]{Rovelli2008}. This prediction can be tested with gamma-ray astronomy (see e.g. \citealp{PhysRevD.80.084017, ELLIS2008412, 2013LRR....16....5A} and references therein). Such theories are most simply expressed in terms of effective field theories, with two consequences. First, general and special relativity are recovered at leading order of the Lagrangian, but higher-order terms, scaling with powers of the energy, appear in the Lagrangian. Second, the effective field theory breaks down at a specific energy, presumably below the Planck scale. These consequences dictate the search strategy and energy scale of interest in the measurements described below. Observations can be used to constrain particular effective field theories (see e.g. \citealp{PhysRevD.92.045016}, where constraints are placed on a specific model, the Standard Model Extension), but the approach taken in most studies has been simpler: to set a lower limit on the energy scale at which a modification to Lorentz invariance can occur. 

\subsection{Theoretical framework}
A natural consequence of quantum gravity is that the uncertainty principle makes the vacuum richer (``fuzzy'' or ``foamy''). This introduces a small refractive index for particles travelling through the vacuum, resulting in an energy-dependent dispersion. 

Rather than the classic relation $E^{2} = p^{2}c^{2}+m^{2}c^{4}$, the energy in effective field theories incorporating LIV can be generically expanded as

\begin{equation}
\label{ch7:eq:energyLIV}
E^{2} = p^{2}c^{2}+m^{2}c^{4} \pm E^{2}\Bigg(\frac{E}{\xi_{n} E_{\rm LIV}}\Bigg)^{n}
\end{equation}
where $c$ is the standard speed of light, $E_{\rm LIV}$ sets the energy scale of the modifications, and the coefficients $\xi_{n}$ give the relative size of the terms in the expansion (note that $\xi_{n}$ can be very large/infinite, rendering terms negligible/zero). The additional terms can be positive or negative, corresponding to a superluminal or subluminal velocity, respectively. The latter is generally considered more theoretically probable \citep{2008PhRvD..78l4010J}. It should be noted that odd-powered terms violate CPT invariance (charge conjugation/parity/time, where parity refers to a spatial inversion), and thus theories that respect CPT will only contain even-powered terms. However, while the standard model of particle physics respects CPT, it can be violated in models incorporating LIV~\cite{2013LRR....16....5A}. 

\subsection{VHE observables}
Gamma rays travelling over large distances will accumulate effects of \gls{liv}, and the agreement between their expected and observed propagation provides a probe of modifications at a given energy scale. This is of particular interest as it is an energy regime that cannot be tested with the particle physics experiments upon which we base our understanding of quantum mechanics and special relativity. 

\subsubsection{Modification of pair-production threshold}
Returning to the earlier discussion of the EBL, the energy threshold for pair production of gamma rays interacting with EBL photons is modified by LIV terms \citep{1999ApJ...518L..21K, 2008PhRvD..78l4010J}. Assuming energy/momentum conservation and considering all particle types to be equally affected by LIV, this modifies Eq.~\eqref{ch7:eq:EBLthreshold} for the subluminal case to 

\begin{equation}
\label{ch7:eq:EBLthresholdLIV}
E_\gamma' \epsilon_{\rm EBL}' \geq \left(m_{\rm e} c^2\right)^2 + \frac{1 - 2^{-n}}{4}\Bigg(\frac{E_\gamma'}{\xi_{n} E_{\rm LIV}}\Bigg)^{n}E_\gamma'^{2}
\end{equation}

\begin{figure}[t]
\centering
\includegraphics[width=0.8\textwidth]{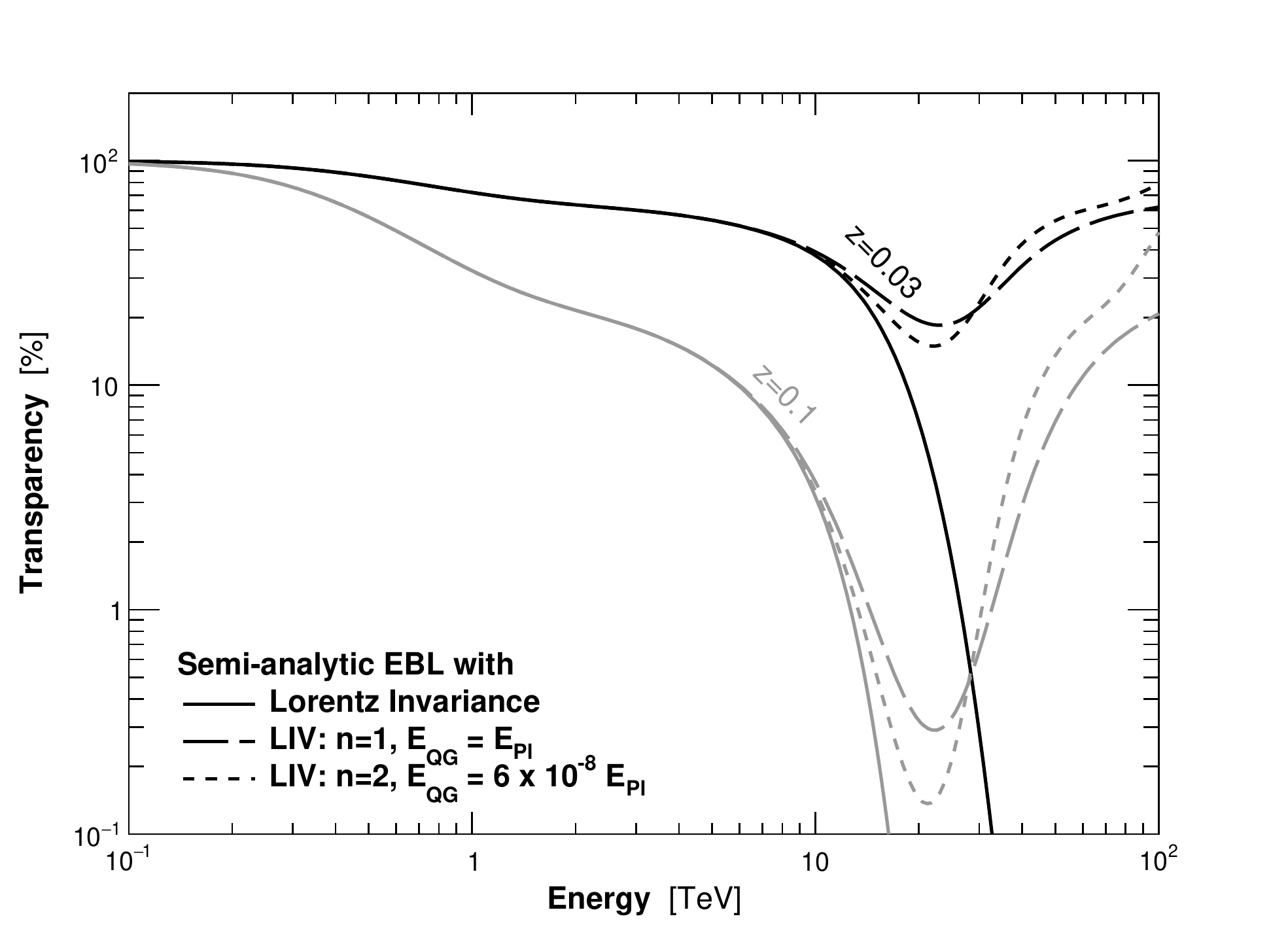}
\caption{Attenuation factor, in percent, as a function of gamma-ray energy on Earth, for sources located at $z=0.03$ and $z=0.1$, following the semi-analytical EBL model of \citet{2012MNRAS.422.3189G} and the LIV formalism of \citet{2008PhRvD..78l4010J}. Both linear ($n=1$) and quadratic ($n=2$) modifications of the pair-creation threshold are shown, at the Planck scale for $n=1$, and at an ad-hoc energy scale of $6\times10^{-8}\:E_{\rm Pl}$ for $n=2$, resulting in a similar effect.}
\label{ch7:fig:EBLopticaldepth_LIV}
\end{figure}

It can be seen from Eq.~\eqref{ch7:eq:EBLthresholdLIV} that LIV raises the energy threshold for pair production for a subluminal modification of the dispersion relation. As shown in Fig.~\ref{ch7:fig:EBLopticaldepth_LIV}, subluminal LIV decreases the opacity of the universe to gamma rays, by effectively depleting the target photon field interacting with gamma rays above 10\:TeV. Assuming that the pair-production cross section retains its functional dependence on energy under LIV, the loss and recovery of transparency are nearly symmetric around a critical energy~\citep{2008PhRvD..78l4010J}, as shown in Fig.~\ref{ch7:fig:EBLopticaldepth_LIV}. As the effect is most noticeable at the highest energies in a gamma-ray spectrum, tests of LIV-induced threshold modifications are best performed using observations of blazars with hard photon spectra extending to tens of TeV.

\subsubsection{Energy-dependent time dispersion}
Another approach to searching for LIV, which requires no assumptions about kinematics, is to search for a correlation between the energy and arrival time of gamma rays. In this case, the limiting assumption is that all of the gamma rays were produced at the same time at the source (we will discuss the uncertainties associated with this assumption below). For an individual photon, a subluminal modification to the velocity is parametrised as

\begin{equation}
\label{ch7:eq:velocityLIV}
v = c\times(1 - (E/\chi_{n} E_{\rm LIV})^{n})
\end{equation}
The parameter $\chi_{n}$ is the equivalent $\xi_{n}$ in Eq.~\eqref{ch7:eq:EBLthresholdLIV} within a numerical factor. A more direct link between Eq.~\eqref{ch7:eq:EBLthresholdLIV} and Eq.~\eqref{ch7:eq:velocityLIV} can be found by defining the velocity following the formalism of Hamiltonian mechanics, that is, as $v = \partial E / \partial p$.

Measuring the arrival time of gamma rays of different energies coming from the same emitter enables the extraction of $E_{\rm LIV}$. The limitation of this method is obvious: how does the observer know that the gamma rays were generated in the source at the same time? Such searches focus on bursts of gamma rays, whether from GRBs, AGN flares, or the well-defined flux peaks in pulsar lightcurves, where a sharp increase in flux occurs across the gamma-ray energy range. For single sources, this method remains vulnerable to intrinsic energy dependence of emission at the source (for instance, due to the presence of different emission regions), which could either increase or decrease the observed time dispersion. 

A robust detection of an LIV effect would consequently rely on the combination of observations from multiple sources, and indeed multiple types of sources, located at different redshifts. The time difference due to LIV should increase as a function of redshift as the effects accumulate over the photon's path length. Detection of an energy-dependent dispersion without such a trend would point to intrinsic source effects. 

In terms of source selection, the properties of GRBs, AGN, and pulsars offer different advantages and disadvantages. The ideal candidates for LIV studies are distant (LIV effects accumulate over distance), rapidly variable (producing well-defined bursts of gamma rays), and detected to high energies (as the modifications to the propagation scale with energy). 

AGN and GRBs are attractive candidates for the first and third properties, located at cosmological distances and with energy spectra that are observed to extend to tens of TeV for AGN and beyond 1\:TeV for GRBs. The flux variability of AGN is however not optimal, as it is unpredictable and often slow (timescales of days to years). GRBs behave more predictably, with a rapid burst of gamma rays followed by a steep decay in the gamma-ray flux, but there is variation between bursts as well as uncertainty in the underlying structure of the flux as a function of time. Pulsars are excellent candidates from the perspective of flux variability. The peaks of their lightcurves are sharp and predictable (enabling improvements in statistical uncertainties of LIV tests by simply observing the sources for longer), and changes in their behaviour, due to e.g.\ spin-down, are precisely predicted. For energy range and distance, though, pulsars are less attractive candidates, located within our galaxy and emitting gamma rays to $\sim$1~TeV rather than tens of TeV. Only four pulsars are detected at VHE; of these, the Crab pulsar is detected to the highest energies, up to $\sim$1.5 TeV \citep{CrabpulsarMAGIC}.

\subsection{VHE constraints}

\subsubsection{Search for modification of pair-production threshold}
The large blazar spectral sample of \citet{2015ApJ...812...60B} facilitated the extraction of limits on $E_{\rm LIV}$, while leaving the EBL SED free within observational constraints. A range of possible $E_{\rm LIV}$ was scanned under the assumption of subluminal LIV, with a recalculation of the optical depth with the modified pair production threshold of Eq.~\eqref{ch7:eq:EBLthresholdLIV}. 
No significant evidence of LIV was observed. Consequently, a lower limit of $E_{\rm LIV}>0.6\times E_{\rm Pl}$ was set at the 99\% confidence level. This can be translated into a limit of $4 \times 10^{-8} \times E_{\rm Pl}$ for a second order modification (see Eq.~(32) in \citealp{2015ApJ...812...60B}).

Similarly, \citealp{PhysRevD.99.043015} evaluated three distinct EBL models and selected the most promising spectra for detecting LIV effects from a sample of 111 archival AGN energy spectra. The final sample included 18 spectra from 6 blazars, dominated by bright and only moderately distant ($z<0.2$) objects. The extracted limits on $E_{\rm LIV}$ for subluminal LIV are well above the Planck energy for the linear ($n=1$) term of Eq.~\eqref{ch7:eq:energyLIV}, with $E_{\rm LIV}<10\times E_{\rm Pl}$ excluded at the 95\% confidence level. For the quadratic term, $E_{\rm LIV}<2\times 10^{-7} \times E_{\rm Pl}$ is excluded.

As for studies of the EBL, analyses exploiting archival data are not devoid of possible biases. In particular for LIV, the treatment of upper limits at the high-energy end of the gamma-ray spectra is expected to impact the constraining power on the LIV energy scale, which could contribute to the differences in the results from the two above-mentioned studies.

\subsubsection{Search for energy-dependent time dispersion}
In 2006, H.E.S.S. observed a massive flare from PKS 2155-304, peaking above 10 times the flux of the Crab Nebula. The source displayed significant variability on the time scale of minutes. A time-dependent analysis was performed wherein the structure of the intrinsic flux variability was assumed to be given by the observed flux variability of the lowest energy gamma rays, between 0.25\:TeV and 0.28\:TeV. No distortion of the lightcurve stage at higher energies was allowed, only an overall shift in arrival time. 

\begin{figure}[t]
\centering
\includegraphics[width=1.0\textwidth]{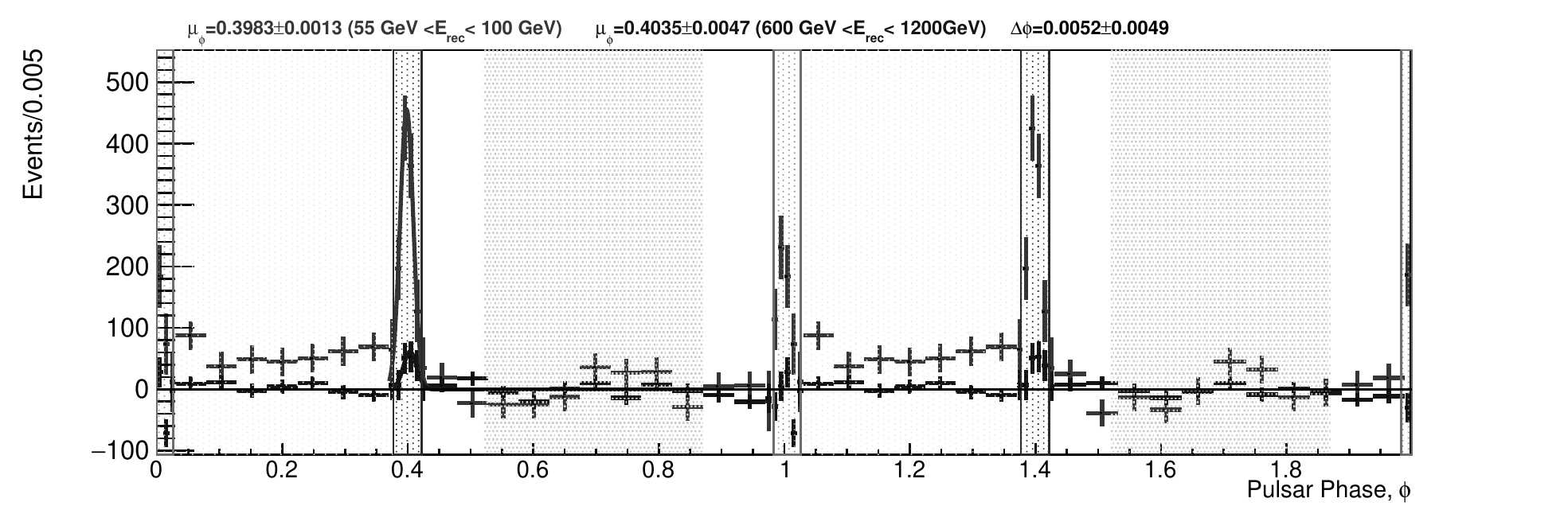}
\includegraphics[width=1.0\textwidth]{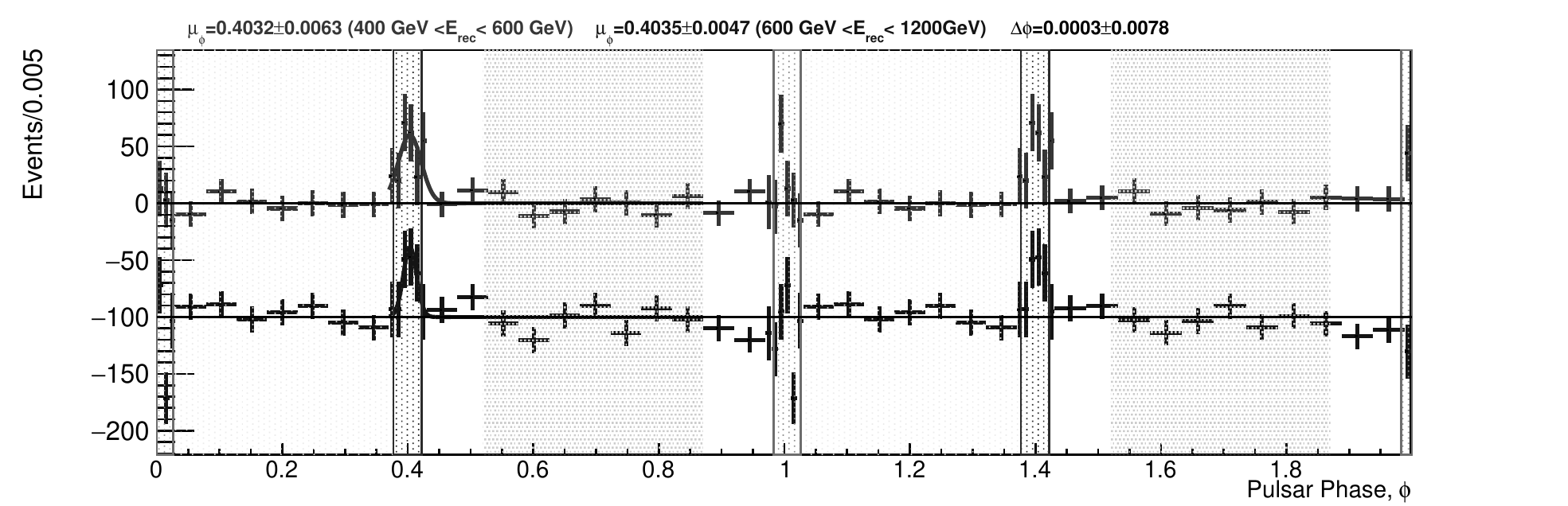}
\caption{The Crab pulsar folded lightcurve as measured by MAGIC. Two periods are shown, for high and low gamma-ray energies. The shaded areas define the pulse and inter-pulse regions. Figure extracted from \citet{MAGICLIVCrabPulsar}.}
\label{ch7:fig:LightcurveLIVCrab}
\end{figure} 

No evidence for an energy-dependent time dispersion was observed, and lower limits were set on $E_{\rm LIV}$: on the linear term at 0.2$\times E_{\rm Pl}$, on the quadratic term at $5\times 10^{-9}\times E_{\rm Pl}$, both at the 95\% confidence level. While these limits were calculated for the subluminal case, the superluminal case gives similar, slightly weaker limits~\citep{2011APh....34..738H}.

A similar study was undertaken by MAGIC for two flares of the blazar Mrk 501, yielding similar limits for the quadratic term to the H.E.S.S. result, and order-of-magnitude weaker limits on the linear term~\citep{ALBERT2008253, MARTINEZ2009226}.

Turning to pulsars, the MAGIC collaboration produced LIV constraints based on 12 years of observations of the Crab pulsar~\citep{MAGICLIVCrabPulsar}. As a pulsar, its flux shows a predictable time structure, in the case of the Crab characterised by two sharp peaks of differing intensity, and a low state or non-detected state in between the peaks, as shown in Fig.~\ref{ch7:fig:LightcurveLIVCrab}. While the variability behaviour of the source is controlled, the distance to the Crab pulsar is to date only measured to 25\% accuracy~\citep{0004-637X-677-2-1201}, which propagates to an uncertainty on the derived LIV results. Both a simple comparison of the peak positions in time for gamma rays in different energy ranges and a full likelihood analysis were performed, yielding consistent results. In the absence of an LIV signal, lower limits were calculated at the 95\% confidence level for $E_{\rm LIV}$, corresponding to $0.05\times E_{\rm Pl}$ for the linear term and  $5\times 10^{-9}\times E_{\rm Pl}$ for the quadratic term, assuming subluminal LIV. As for the results of \citet{2011APh....34..738H}, the limits for the superluminal case are slightly weaker.

With the detection of the first GRBs at VHE, a new source class is available for time dispersion measurements. In particular, the MAGIC-detected GRB190114C had the large redshift ($z\sim0.4$) and high-energy photons (largest detected photon energy was ${\sim}\,2\:$TeV) necessary to produce competitive LIV constraints~\citep{PhysRevLett.125.021301}. The dominant uncertainty in this measurement is the shape of the intrinsic emission as a function of time. The authors considered a minimal model, which assumes zero emission before the multiwavelength GRB alert was issued, triggering MAGIC observations, as well as a theoretically motivated model in which the peak emission occurs before the beginning of MAGIC observations. Neither model of the intrinsic emission provided evidence for LIV, and the extracted 95\% confidence level limits were comparable, with $0.5\times E_{\rm Pl}$ for the subluminal linear term, assuming the physically motivated emission model, and $5\times 10^{-9}\times E_{\rm Pl}$ for the subluminal quadratic term. Such limits suggest a sensitivity similar to that of AGN time dispersion measurements: the lower energy of gamma rays from the GRB are compensated by the greater redshift of the source.

\begin{figure}[t]
\centering
\includegraphics[width=1.0\textwidth]{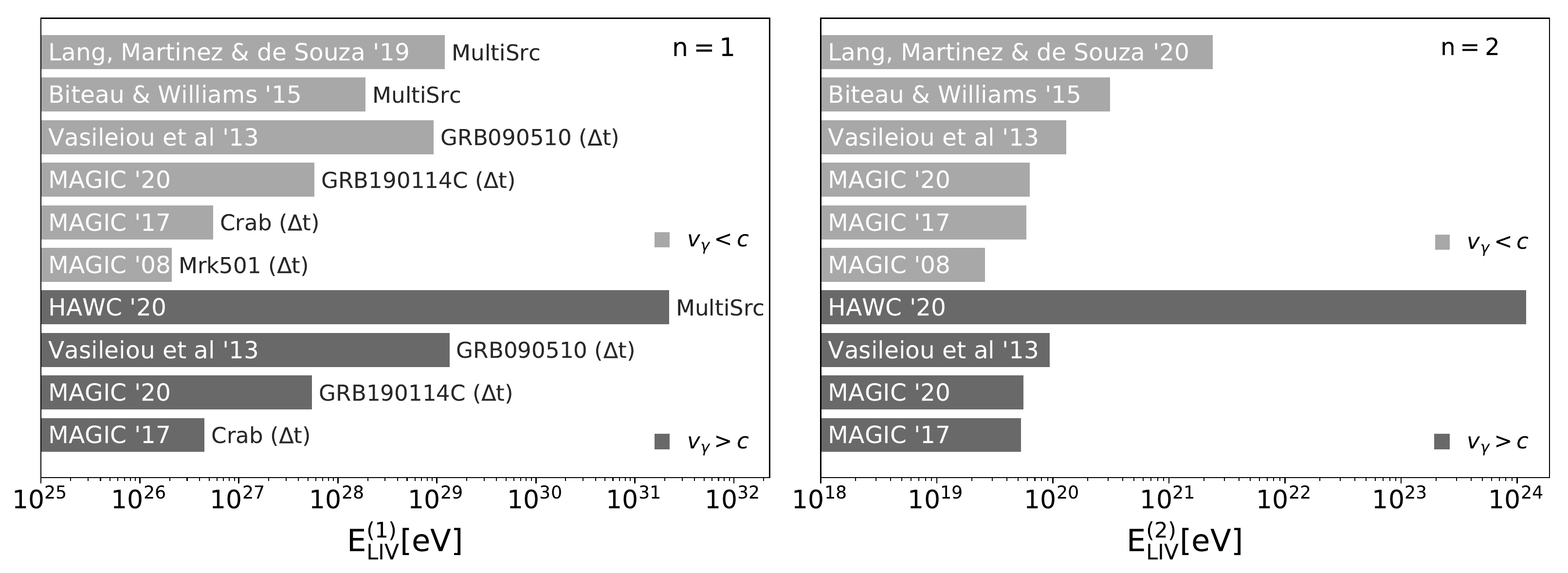}
\caption{Summary of the lower limits on the energy scale of LIV, $E_{LIV}$. The left panel shows the limits on the linear term, the right panel on the quadratic term. Light grey bars assume a subluminal modification, while dark grey bars assume a superluminal modification. Credit: Humberto Mart{\'{\i}}nez-Huerta, adapted from \citet{sym12081232}.}
\label{ch7:fig:LIVlimits}
\end{figure}

A summary of the limits discussed here, including the non-IACT measurements discussed in the next section, is given in Fig~\ref{ch7:fig:LIVlimits}. Further discussion of current limits on LIV can be found in \citet{sym12081232}.

\subsection{Relation to non-\textbf{IACT} measurements}
The most constraining limits on the subluminal linear term for LIV come from \LAT\. These constraints are based on observations of GRBs, which within the \LAT\ sensitive energy range are ideal candidates, with sharp flux variability and broadband emission, and occurring at cosmological distances. A large sample of GRBs has been measured, although several of the canonical limits are from individual \citep{FermiLimits} or small samples of GRBs \citep{PhysRevD.87.122001}. However, GRBs exhibit intrinsic energy-dependent time lags which must be taken into account \citep{0067-0049-209-1-11}. Lower limits at the 95\% confidence level on $E_{\rm LIV}$ exceed the Planck scale by a factor of 7.6 for the linear term, going down to a factor of 2 when accounting for systematic uncertainties related to emission processes at the source. For the quadratic term, the limits are comparable to many of the VHE limits, at $1\times 10^{-8}\times E_{\rm Pl}$~\citep{PhysRevD.87.122001}.

The HAWC observatory has strongly constrained superluminal LIV by searching for high-energy spectral cutoffs induced by photon decay to an electron/positron pair or to multiple photons~\citep{PhysRevLett.124.131101}. This search utilised four Galactic high-energy emitters, including the Crab Nebula, and placed 95\% confidence level lower limits on $E_{\rm LIV}$ three orders of magnitude above the Planck energy for the linear term and $7\times 10^{-6} \times E_{\rm Pl}$ on the quadratic term.

Upper limits on the flux of photons at ultra-high energies ($E>10^{18}\:$eV) might also provide stringent constraints on LIV, provided the sources are identified and a significant fraction of protons is present at the highest energies. Such a probe could reach the Planck scale even for second-order modification of the pair-creation threshold \citep{Gelmini2008, 0004-637X-853-1-23}. LIV could also impact the spectrum of UHECR by increasing the threshold for photo-production of pions (see Sec.~3.5 in \citealp{{2013LRR....16....5A}} for an excellent review of the debates on this matter). The constraints on the composition and sources of UHECR expected from the upgrade of the Pierre Auger Observatory will help in determining the magnitude of the constraints on LIV from UHECR propagation.

\subsection{Outlook}

The violation of Lorentz invariance at energies beyond the reach of man-made accelerators can be studied with gamma rays at GeV-TeV energies. Two paths have been followed in the VHE regime, both based on phenomenological modifications of the dispersion relation of photons, resulting in high-energy correction terms.

Assuming that speed can be computed from Hamiltonian mechanics within a quantum-gravity framework, such modifications result in an energy-dependent speed of light, whose effect, cumulated over cosmological scales, could be observed as an energy-dependent time-lag between successive gamma-ray bands. The lack of such lags has already allowed ground-based instruments to place limits on the LIV energy scale that, for a linear modification, approach the Planck scale. 
The timing capabilities of CTA will be more than three orders of magnitude better than {\it Fermi}-LAT in the 30-100\:GeV band \citep{2013APh....43..348F}. This will facilitate the detection of numerous highly variable extragalactic sources (GRBs, blazars), constituting a gold mine for LIV studies.

An alternative approach, based on four-momentum conservation, suggests that LIV corrections should result in a modification of the pair-creation threshold, directly affecting the transparency of the universe beyond tens of TeV. In contrast to timing approaches, the limits for a linear modification have been quoted an order of magnitude above the Planck scale. However, limits on the subluminal quadratic modification, which is particularly motivated from the theoretical perspective, remain well below the Planck energy, on the order of $10^{-7}\times E_{\rm Pl}$. Thanks to its enhanced sensitivity at the highest energies, CTA will be able to probe LIV predictions in both nearby \citep{2014JCAP...06..005F} and more distant objects \citep{2016A&A...585A..25T}, either improving existing constraints or finding evidence for LIV \citep{2021JCAP...02..048A}.

The two approaches, based on timing on the one hand and spectral reconstruction on the other, are complementary in their assumptions: the first assumes an energy-dependent modification in propagation time, the second a modification to the kinematics of pair production. The discovery of LIV in one channel does not demand it in the other. The surprise with CTA may come from studies of the second-order modification, which affects gamma-ray observables at an LIV energy scale beyond $10^{20}\:$eV.

\section{Status and prospects of TeV gamma-ray cosmology}
In this chapter, we have attempted to demonstrate the relevance of VHE gamma-ray observations for understanding cosmology. We have examined time scales ranging from the Planck time to the current day, considering Lorentz invariance violation and cosmological magnetic and photon fields. The current generation of IACTs has made progress in measuring or constraining all of these cosmological observables. We summarise by pointing out that gamma-ray-based measurements are beginning to probe EBL wavelengths where theoretical models developed prior to the gamma-ray detections disagree, are rapidly narrowing the allowed range of intergalactic magnetic field strength and correlation length, and are pushing the lower limit on the energy scale at which Lorentz invariance violation may occur beyond the Planck scale. We have also discussed the relevance of these measurements for other fields of astrophysics and astroparticle physics, particularly the spectral energy distribution of the diffuse supernova neutrino background and the propagation of UHECR. We have noted the extent to which the gamma-ray measurements are limited by astrophysical uncertainties, namely those associated with the production of VHE gamma rays in astrophysical sources. While current-generation IACTs have not made their final statements on the topics discussed above, we can look to CTA to substantially improve our knowledge and possibly make ground-breaking discoveries. The precise energy and angular resolution of CTA, coupled with its high flux sensitivity, will allow it to quickly supersede existing instruments, dramatically improving our understanding of the light content of the universe, cosmic magnetism, and the structure of space time.

\appendix
\newpage
\section*{Exercise 1}

The energy density of the CMB at the present epoch is about $\rho_{\rm CMB} \sim 0.26$\:eV\:cm$^{-3}$ and the integral cross section for pair production can be estimated to be ${\sim}\sigma_{\rm T}/10$, where $\sigma_{\rm T}$ is the Thomson cross section. Gamma rays with energies around 1\:TeV mostly interact with photons from COB. 
\begin{enumerate}
\item Estimate the energy density of COB photons, using the information given in the text about the relative budgets of the CMB and EBL. 
\item Modeling the EBL spectrum as a Dirac function peaked at 1\:eV, and noting that 1\:TeV gamma rays pair produce mainly with 1\:eV photons, compute the optical depth, $\tau$, for 1\:TeV gamma rays as a function of distance $L \sim cz/H_0$ (valid for $z\ll1$) in Gpc.
\item Where is the cosmic gamma-ray horizon, defined as the distance for which $\tau=1$, located for 1\:TeV gamma rays?
\end{enumerate}
\textit{Note: The cosmic gamma-ray horizon for 1\:TeV gamma rays is typically reached for sources at $z\sim0.1$, check if the value of the corresponding luminosity distance is consistent with your rough estimate.}\\

\noindent{\it- List of constants -}
\begin{itemize}
\item$\sigma_{\rm T} \sim 6.6 \times 10^{-25}$\:cm$^{2}$
\item1\:pc $ \sim 3.1 \times 10^{18}$\:cm
\end{itemize}

\section*{Exercise 2}

The total pair-production cross section, under the assumption of an isotropic distribution of EBL photons, can be approximated by \citep{1990MNRAS.245..453C}:
\begin{equation}
\sigma(x) = \sigma_0 \times \frac{x^2-1}{x^3} \ln x \quad {\rm for\ }x>1
\end{equation}
where $x = E_\gamma \epsilon_{\rm EBL}$ is the product of the energies of the gamma ray and of the EBL photon.

\begin{enumerate}
\item Locate numerically the maximum of the total pair-production cross section.
\item Compute the wavelength of EBL photons which are most likely to interact with 0.1, 1, and 10\:TeV gamma rays.
\end{enumerate}

\noindent{\it- List of constants -}
\begin{itemize}
\item$\hbar c \sim 197$\:MeV\:fm
\end{itemize}

\section*{Exercise 3}

Electrons and positrons produced by the interaction of gamma rays with EBL photons can inverse Compton scatter on CMB photons, boosting them to higher energies. In the Thomson regime (the low-energy limit of Compton scattering), the average energy of the secondary gamma rays is:
\begin{equation}
\label{ch7:eq:Thomson}
E_{\gamma} = \frac{4}{3} \gamma_{\rm e}^2 \epsilon_{\rm CMB}
\end{equation}
where $\epsilon_{\rm CMB}\sim 0.6$\:meV is the average CMB photon energy, and where $\gamma_{\rm e}$ is the Lorentz factor of the electron/positron.

\begin{enumerate}
\item The Thomson regime is defined as $\gamma_{\rm e} \epsilon_{\rm CMB} \ll m_{\rm e} c^2$. Up to which electron Lorentz factor is Eq.~\eqref{ch7:eq:Thomson} valid?
\item Compute the energy of the secondary gamma rays for primary photons at 0.2, 2, and 20\:TeV.
\end{enumerate}

\section*{Exercise 4}

Phenomenological modifications to the dispersion relation of particles at high energies can be written as :

\begin{equation}
E^{2} = p^{2}c^{2}+m^{2}c^{4} \pm E^{2}\Bigg(\frac{E}{\xi_{n} E_{\rm LIV}}\Bigg)^{n}
\end{equation}
where $E_{\rm LIV}$ is the energy scale at which Lorentz invariance violation appears and $n$ is the order of the correction. In a Hamiltonian formalism where $v = \partial E / \partial p$, this results in subluminal/superluminal motion of particles.

\begin{enumerate}
\item Show that to order $n$, for $m \ll E \ll E_{\rm LIV}$ : $v = c\times\left(1 \pm (E/\chi_{n} E_{\rm LIV})^{n}\right)$, with $\chi_{n} = \left(\frac{n+1}{2}\right)^{-n}\xi_{n}$
\item Compare the relative size of the coefficients giving the LIV scale obtained from time delays and threshold effects for first and second-order modifications of the dispersion relation.
\end{enumerate}

\bibliographystyle{abbrvnat}
\bibliography{biblio/bib_HESS,biblio/bib_VERITAS,biblio/bib_MAGIC,biblio/bib_LAT,biblio/bib_coll,biblio/bib_ch7}

\end{document}